\documentclass{aastex61}
\usepackage{graphicx,natbib}
\usepackage[percent]{overpic}
\usepackage{graphicx,subfigure,amsmath}
\newcommand\gsim{\,\lower3pt\hbox{$\sim$}\llap{\raise2pt\hbox{$>$}}\,}
\newcommand\lsim{\,\lower3pt\hbox{$\sim$}\llap{\raise2pt\hbox{$<$}}\,}

\shorttitle{eruption of flux ropes}
\shortauthors{Fan}

\begin{document}

\title{MHD simulations of the eruption of coronal flux ropes under
coronal streamers}

\correspondingauthor{Yuhong Fan}
\email{yfan@ucar.edu}

\author{Yuhong Fan}
\affil{High Altitude Observatory, National Center for Atmospheric Research, \\
3080 Center Green Drive, Boulder, CO 80301, USA}

\begin{abstract}
Using three-dimensional magnetohydrodynamic (MHD) simulations, we
investigate the eruption of coronal flux ropes underlying coronal streamers
and the development of a prominence eruption.
We initialize a quasi-steady solution of a coronal helmet
streamer, into which we impose at the lower boundary the slow
emergence of a part of a twisted magnetic torus.
As a result a quasi-equilibrium flux rope is built up under the streamer.
With varying sizes of the streamer and the different
length and total twist of the emerged flux rope, we found different scenarios
for the evolution from quasi-equilibrium to eruption. In the cases with a
broad streamer, the flux rope remains
well confined until there is sufficient twist such that it first develops the
kink instability and evolves through a sequence of kinked, confined states
with increasing height until it eventually develops a ``hernia-like'' ejective
eruption.  For the significantly twisted flux ropes,
prominence condensations form in the dips of the twisted field lines due to
run-away radiative cooling.
Once formed, the prominence carrying field becomes significantly non-force-free
due to the prominence weight despite being low plasma $\beta$.
As the flux rope erupts, we obtain the eruption of the prominence, which shows
substantial draining along the legs of the erupting flux rope.
The prominence may not show a kinked morphology even though the flux rope
becomes kinked.
On the other hand, in the case with a narrow streamer, the flux
rope with less than 1 wind of twist can erupt via the onset of the torus instability.
\end{abstract}


\section{Introduction}
Solar prominences or filaments (elongated large-scale structures of cool and
dense plasma suspended in the much hotter and rarefied solar corona) are
major precursors or source regions of coronal mass ejections (CMEs).
It is suggested that most CMEs are the result of the destabilization and
eruption of the hosting magnetic structure capable of supporting the
prominence \citep[e.g.][]{Webb:Hundhausen:1987}.
The hosting magnetic structure is likely a magnetic flux rope with helical
field lines twisting about its center, supporting the dense prominence plasma
at the dips of the field lines \citep[e.g.][]{Low:2001,Gibson:2015}.
Many previous MHD simulations of CME initiation have focused on the mechanism
for the destabilization and eruption of coronal flux ropes using highly
simplified thermodynamics, e.g. zero-plasma $\beta$, isothermal, or an ideal
gas with lowered adiabatic index $\gamma$, without the possible formation
and the effects of the prominence condensations \citep[e.g.][]{Antiochos:etal:1999,
Toeroek:Kliem:2005,Toeroek:Kliem:2007,Toeroek:etal:2011,Aulanier:etal:2010,
Fan:Gibson:2007,Fan:2010,Fan:2012, Chatterjee:Fan:2013,Amari:etal:2014}.
These simulations also generally
consider an initial static potential field without an ambient solar wind
that partially opens the magnetic field.
MHD simulations of CME events with more realistic treatment of the
thermodynamics that explicitly include the non-adiabatic effects of the
corona and transition region, termed
the thermodynamic MHD simulations \citep[e.g.][]{Linker:etal:2007,Downs:etal:2012},
have been conducted.
These simulations have been used to carry out direct
comparison with coronal multi-wavelength
observations of the simulated events. For example, \citet{Downs:etal:2012}
have conducted a global thermodynamic MHD simulation of the 2010 June 13
CME event and used forward modeled EUV observables to compare and
interpret SDO/AIA observations of the EUV
waves associated with the eruption. However these studies have not
presented explicitly the formation and eruption of prominences.
Recently, \citet{Xia:etal:2014,Xia:Keppens:2016a} have carried out
thermodynamic MHD simulations to model the formation and fine-scale
dynamics of a prominence in a stable equilibrium coronal
flux rope, which reproduced many observed features of solar prominences by SDO/AIA. 
However, simulations of the eruption of prominence-carrying coronal flux ropes
are still an area to be explored.
In this paper we carry out MHD simulations of the evolution of coronal flux
ropes under coronal streamers, and explicitly include the non-adiabatic
effects that allow for the formation of prominence condensations, and model
the destabilization and eruption of the flux ropes with the more realistic
treatment of the thermodynamics.
We assume a fully ionized hydrogen gas with the
adiabatic index $\gamma = 5/3$, and explicitly include a simple empirical coronal
heating, optically thin radiative losses, and the field aligned thermal conduction.
We consider a broad and a narrow initial coronal streamer into which we drive
the emergence of a twisted flux rope with a varying length and total twist, and
we find different scenarios and mechanisms for the transition from quasi-equilibrium
to dynamic eruption of the flux rope.
In the cases with a long, significantly twisted flux rope, we also find the
formation of prominence condensations in the dips of the twisted field lines
due to the development of the radiative instability or non-equilibrium.
With the eruption of the flux rope, we also obtain a prominence eruption.
\citet{Pagano:etal:2014} and \citet{Pagano:etal:2015a} have also carried out
MHD simulations of flux rope ejection incorporating field aligned thermal
conduction and optically thin radiative losses, and have synthesized the
modeled SDO/AIA EUV emissions from the simulated eruption.
Their thermodynamic MHD simulations focus on the dynamic eruption phase,
using an initial flux rope configuration that is already out of equilibrium
as evolved from a separate zero plasma-$\beta$ global magnetofriction model
\citep{Mackay:vanB:2006b}.
Our MHD simulations on the other hand also model the transition from
quasi-equilibrium to the development of the instabilities of the flux rope.

\section{The Numerical Model}
\begin{figure}[htb!]
\centering
\includegraphics[width=0.5\textwidth]{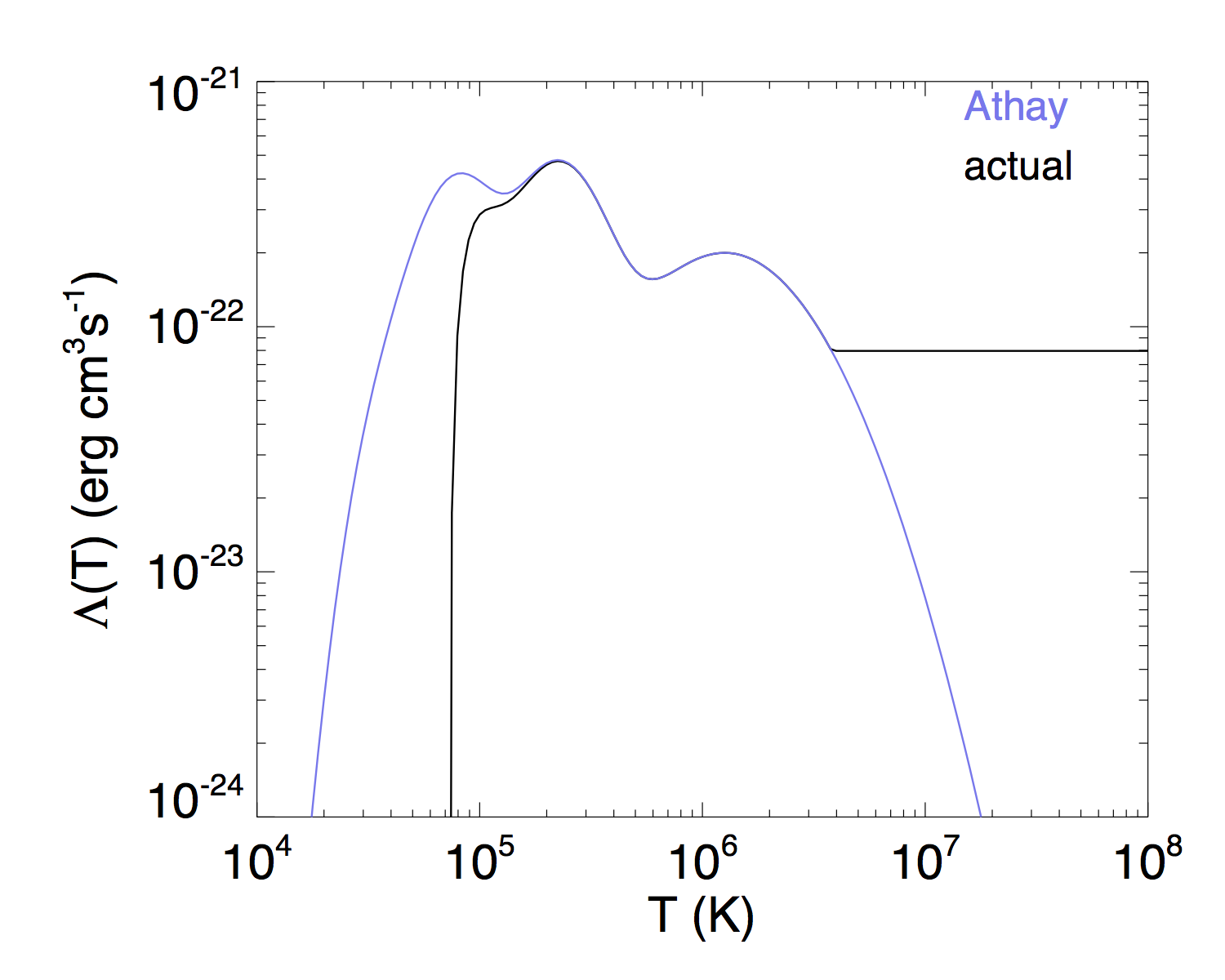}
\caption{The radiative loss function as given in \citet{Athay:1986}
(blue curve), and the modified actual function used, where the
radiative loss is suppressed for $T \leq 7 \times 10^4$ K
and is set to constant for $T$ above $3.8 \times 10^6$ K.}
\label{fig:radloss}
\end{figure}

For the simulations, we solve the following semi-relativistic MHD equations
\citep{Gombosi:etal:2002, Rempel:2017} in spherical geometry:

\begin{equation}
\frac{\partial \rho}{\partial t}
= - \nabla \cdot ( \rho {\bf v}) ,
\label{eq:cont}
\end{equation}
\begin{eqnarray}
\frac{\partial ( {\rho \bf v})}{\partial t}
&=& - \nabla \cdot \left ( \rho {\bf v} {\bf v} \right )
- \nabla p + \rho {\bf g} + \frac{1}{4 \pi}
( \nabla \times {\bf B} ) \times {\bf B} 
\nonumber \\
& & + \frac{{v_A}^2/c^2}{1 + {v_A}^2/c^2} \left[ {\cal I}
- {\hat{\bf b}} {\hat{\bf b}} \right ]
\cdot \left [ ( \rho {\bf v} \cdot \nabla ) {\bf v} + \nabla p - \rho {\bf g}
- \frac{1}{4 \pi} ( \nabla \times {\bf B} ) \times {\bf B} \right ] ,
\label{eq:motion}
\end{eqnarray}
\begin{equation}
\frac{\partial {\bf B}}{\partial t}
= \nabla \times ({\bf v} \times {\bf B}),
\label{eq:induc}
\end{equation}
\begin{equation}
\nabla \cdot {\bf B} = 0,
\label{eq:divb}
\end{equation}
\begin{equation}
\frac{\partial e}{\partial t} = - \nabla \cdot
\left ( {\bf v} e \right ) - p \nabla \cdot {\bf v}
- \nabla \cdot {\bf q} + Q_{\rm rad} + H  ,
\label{eq:energy}
\end{equation}
\begin{equation}
p = \frac{\rho R T}{\mu},
\label{eq:state}
\end{equation}
where
\begin{equation}
e = {\frac{p}{\gamma - 1} } ,
\label{eq:eint}
\end{equation}
\begin{equation}
v_A = \frac{B}{\sqrt {4 \pi \rho}} ,
\label{eq:alfvenspeed}
\end{equation}
and the last term on the right hand side of the
momentum equation (eq. (\ref{eq:motion})) is the 
semi-relativisitc correction (see eqs. (53) (54) in
\citet{Rempel:2017}).
In the above, symbols have their usual meanings,
where ${\bf v}$ is the velocity field, ${\bf B}$ is the magnetic field,
$\rho$, $p$, and $T$ are respectively the plasma density, pressure and
temperature, $e$ is the internal energy density, $c$ is the
(reduced) speed of light (see more discussion later),
${\cal I}$ is the unit tensor, ${\hat {\bf  b}} = {\bf B} / B$ is the unit
vector in the magnetic field direction,
$R$, $\mu$, and $\gamma$, are respectively
the gas constant, the mean molecular weight, and the adiabatic index of
the perfect gas, and ${\bf g} = - ( G M_{\odot} / r^2 ) {\bf r}$ is the
gravitational acceleration, with $r$ being the radial distance to the center
of the Sun. We assume a fully ionized hydrogen gas with the adiabatic index
${\gamma} = 5/3$.  We solve the internal energy equation,
taking into account the non-adiabatic effects of an
empirical coronal heating $H$ (to heat the corona and accelerate the
solar wind), optically thin radiative cooling $Q_{\rm rad}$, and electron
heat conduction $- \nabla \cdot {\bf q}$. The conduction
heat flux ${\bf q}$ is given by
\begin{equation}
{\bf q} = f_s {\bf q}_S + (1-f_s) {\bf q}_H ,
\label{conductflux}
\end{equation}
which combines the collisional form of Spitzer:
\begin{equation}
{\bf q}_s=-\kappa_0 T^{5/2} \hat{\bf b} \hat{\bf b} \cdot \nabla T
\label{eq:spitzer}
\end{equation}
with $\kappa_0=10^{-6} {\rm erg}
\; {\rm s}^{-1} \; {\rm cm}^{-1} \; {\rm K}^{-7/2}$,
and the collisionless form given by \citet{Hollweg:1978}:
\begin{equation}
{\bf q}_H = \frac{3}{4} \alpha p {\bf v} ,
\label{eq:hollweg}
\end{equation}
where $\alpha=1.05$,
using an $r$ dependent weighting function
\begin{equation}
f_s = \frac{1}{1+(r/r_H )^2},
\end{equation}
where $r_H = 5 R_{\odot}$, such that the heat flux transitions smoothly
from the collisional form in the lower corona to the collisionless form
at large distances (with $r > 5 R_{\odot}$).
This formulation for ${\bf q}$ is the same as that used in
\citet{vanderHolst:etal:2014}.
The optically thin radiative cooling is given by:
\begin{equation}
Q_{\rm rad} = N^2 \Lambda (T) ,
\label{eq:radloss}
\end{equation}
where $N = \rho / m_p$ is the proton number density assuming a fully ionized
hydrogen gas, with $m_p$ being the proton mass, and the radiative loss function
$\Lambda (T)$ is as given in \citet{Athay:1986}, modified to suppress
cooling for $T \leq 7 \times 10^4$ K and set to constant for
$T > 3.8 \times 10^6$ K as shown in Figure \ref{fig:radloss}.
We suppress cooling for $T \leq 7 \times 10^4$ K so that the smallest pressure
scale height of the coolest plasma that can form does not go below two
grid points given our simulation resolution.
We set the cooling to constant for $T > 3.8 \times 10^6$ K
so it follows more closely to the radiative loss given by CHIANTI 7 with
coronal abundances \citep{Landi:etal:2012} in the high temperature range
near $10^7$ K.
We use a simple form of empirical coronal heating function
that only varies with height following an exponential decay:
\begin{equation}
H = \frac{F}{L_H} \frac{R^2_{\odot}}{r^2}
\exp \left[ -(r - R_{\odot}) / L_H \right ] .
\label{eq:heatingfunc}
\end{equation}
where the input energy flux density is
$F=9.74 \times 10^5 \, {\rm ergs} \, {\rm cm}^{-2} {\rm s}^{-1}$,
the decay length is $L_H = 1.948 \times 10^{10}$ cm.

The above MHD equations are solved using the ``Magnetic Flux Eruption'' (MFE)
code that has been used in several past simulations of coronal mass ejections
\citep[e.g.][]{Fan:2012, Fan:2016}. The code uses a staggered $r$-$\theta$-$\phi$
grid with a second-order accurate spatial discretization.  A second-order
TVD Lax-Friedrichs scheme with a reduced numerical diffusive flux
\citep[eq. A(3) in][]{Rempel:etal:2009} is used for evaluating the advection
terms in the continuity, momentum, and the internal energy equations.
For the induction
equation, a method of characteristics that is upwind for the shear Alfv\'en
waves is applied for evaluating the electromotive force (emf)
${\bf v} \times {\bf B}$
and the constrained transport scheme is used to ensure
$\nabla \cdot {\bf B} = 0 $ \citep{Stone:Norman:1992:b}.
No explicit viscosity and magnetic diffusion are included in the momentum
and the induction equations. However, numerical diffusions are present as a result
of the upwinded evaluations of the fluxes in the momentum equation and the
electric field in the induction equation. 
In the numerical code, the dissipation of kinetic and magnetic energies due
to these numerical diffusive fluxes are being evaluated and added to the
internal energy as non-adiabatic heating, in addition to the explicit
empirical coronal heating term $H$ in the above internal energy equation.
To evaluate the numerical dissipation of magnetic energy, we take the
difference between the actual (upwinded) evaluation of the emf
$( {\bf v} \times {\bf B} )_{\rm actual}$
and a direct linear interpolation evaluation of the emf
$( {\bf v} \times {\bf B} )_{\rm int}$ at the cell edges (where the emf is
defined on the staggered grid). From this difference emf ${\bf E}_ {\rm num}
=( {\bf v} \times {\bf B} )_{\rm actual} - ( {\bf v} \times {\bf B} )_{\rm int}$
we evaluate the conversion of magnetic energy into thermal energy due to
numerical diffusion as
$- (1/4 \pi) {\bf E}_{\rm num} \cdot (\nabla \times {\bf B}) $, which is
added to the right hand side of the internal energy equation.
Here $\nabla \times {\bf B}$ is computed with a simple centered finite
difference.
Similarly, in the evaluation of the advection term
$\nabla \cdot (\rho {\bf v} {\bf v})$
of the momentum equation,
the actual upwinded evaluation of the momentum
flux $(\rho {\bf v} {\bf v})_{\rm actual}$ can be written as
$(\rho {\bf v} {\bf v})_{\rm int} + {\cal D}_{\rm num}$ where
$(\rho {\bf v} {\bf v})_{\rm int}$ is
a direct linear interpolation evaluation
of the momentum flux and ${\cal D}_{\rm num}$ denotes the
additional numerical diffusive flux due to the upwinding in
the modified TVD Lax-Friedrichs scheme.
Then we evaluate $({\cal D}_{\rm num} \cdot \nabla ) \cdot {\bf v}$ as the numerical
viscous dissipation added to the right hand side of
the internal energy equation, where the derivatives for the
$\nabla$ in the above expression is evaluated using simple
centered finite differences.
The (numerical) energy dissipation is the strongest at current sheets and
shocks and is being put back into the internal energy as heating. 

The code uses an explicit 3rd order Runge-Kutta scheme
for temporal discretization.
Under typical coronal conditions, the CFL time step required for the
parabolic (field-aligned) heat conduction is often orders of
magnitude smaller than the time step required for the hyperbolic advection
terms. Here we have used the hyperbolic heat conduction approach
described in section 2.2 in \citet{Rempel:2017} to circumvent the stringent
timestep constraint.
Instead of using equation \ref{eq:spitzer}, we solve the following equation
for the heat flux ${\bf q}_s$:
\begin{equation}
\tau_s \frac{ \partial {\bf q}_s}{\partial t}
=\left ( -\kappa_0 T^{5/2} \hat{\bf b} \hat{\bf b} \cdot \nabla T
-{\bf q}_s \right ),
\label{eq:hpbcond}
\end{equation}
where $\tau_s$ represents a finite time scale
for ${\bf q}_s$ to evolve towards the Spitzer heat flux, and it is
set to \begin{equation}
\tau_s = \frac{\kappa_0 T^{7/2}}{e} \frac{(\Delta t)^2}
{(f_{\rm cfl} \Delta x_{\rm min})^2},
\label{eq:tau_s}
\end{equation}
where $\Delta t$ denotes the dynamic CFL time step determined from
all of the other advection terms, $f_{\rm cfl}$ is the CFL number
used, and $\Delta x_{\rm min}$ is the smallest grid size.
With the addition of equation \ref{eq:hpbcond}, we have 3 more dependent
variables (3 vector component of ${\bf q}_s$) which we advance using the same
Runge-Kutta scheme with the right hand side of equation \ref{eq:hpbcond}
evaluated with a simple 2nd order finite difference scheme on the staggered
grid (without the need for upwinded interpolation).
As described in \citet{Rempel:2017}, the introduction of ${\bf q}_s$
given by equation \ref{eq:hpbcond} produces a wave-like
equation for $T$ (or internal energy $e$), and the above
specification of $\tau_s$ by equation \ref{eq:tau_s} ensures that
the effective wave speed does not require a CFL time step that is
below $\Delta t$.
\citet{Rempel:2017} showed that the hyperbolic heat conduction
approach gives a good approximation of the evolution produced
by the parabolic heat conduction if the required $\tau_s$ is small
compared to the thermal diffusion time scale of interest.
For our current simulations, we have tested this hyperbolic heat
conduction approach by comparing the results obtained using this approach
to those produced by solving for the parabolic heat conduction term
using operator split and sub-cycling with the explicit Super TimeStepping
scheme of \citet{Meyer:etal:2012}, and we found good agreement in
the resulting evolution.

Our simulations are carried out in a spherical wedge domain with
$r \in [R_{\odot}, 11.47 R_{\odot}]$,
$\theta \in [75^{\circ}, 105^{\circ}]$,
and $\phi \in [-\phi_{\rm max},\phi_{\rm max}]$, where we have run
cases with $\phi_{\rm max} = 75^{\circ}$ and $\phi_{\rm max} = 37.5^{\circ}$,
to accommodate flux ropes with different
lengths and total twists.
We use a grid of $504(r) \times 196 (\theta) \times 960 (\phi)$
for the longer domain with $\phi_{\rm max} = 75^{\circ}$
and a grid of $504(r) \times 196 (\theta) \times 480 (\phi)$ for the
domain with $\phi_{\rm max} = 37.5^{\circ}$.
The grid is uniform in $\theta$ and $\phi$ and stretched in the
$r$ direction, with a grid size of $dr = 0.002727 R_{\odot} = 1.898 $ Mm
for $r < 1.79 R_{\odot}$ and it increases geometrically to about
$0.19 R_{\odot}$ at the outer boundary of $r = 11.47 R_{\odot}$.

For the thermodynamic conditions at the lower boundary of the
simulation domain, we assume a fixed transition region temperature of
$5 \times 10^5$ K, and adjust the base pressure $p_{R_{\odot}}$ in the
following way:
\begin{equation}
\frac{ \partial p_{R_{\odot}}}{\partial t}
= \frac{1}{\tau} \left ( p_{\rm reb} - p_{R_{\odot}} \right )
\label{eq:lbpressure}
\end{equation}
where,
\begin{equation}
p_{\rm reb} = C \left ( \kappa_0 T^{5/2} \frac{dT}{dr} 
\right )_{r={R_{\odot}}}
\label{eq:p_reb}
\end{equation}
such that the base pressure is driven towards a value $p_{\rm reb}$ that
is proportional to the downward heat conduction flux, in a time scale of
$\tau$, to crudely represent the effect of chromospheric evaporation.
The expression (eq. [\ref{eq:p_reb}]) for the coronal base pressure
$p_{\rm reb}$ is given
by the radiative energy balance (reb) model of \citet{Withbroe:1988}.
Here we have used $C = 1.32 \times 10^{-6}$ in CGS units and
$\tau = 357 $ sec. We have chosen $\tau$ to be on the order of
the sound crossing time of the chromosphere.
Thus the (time-varying) pressure $p_{R_{\odot}}$ prescribed at the
bottom boundary
via equation \ref{eq:lbpressure} provides a mass reservoir for
the corona plasma and the solar wind that is heated and accelerated in
the domain. Note also this mass reservoir at the lower boundary is
fixed at the temperature of $5 \times 10^5$ K,
so any cool prominence condensation that develops in the corona is
not directly carried into the domain from the lower boundary, but forms after
emerging into the coronal domain.

At the lower boundary we also impose a kinematic magnetic flux transport
by specifying a time-dependent transverse electric field
${\bf E}_{\perp}|_{r=R_{\odot}}$.
Setting the electric field to zero would correspond to a rigid
anchoring or line-tying lower boundary.
At certain time periods during the simulations, we impose the emergence
of a twisted magnetic torus at the lower boundary by specifying
the electric field:
\begin{equation}
{\bf E}_{\perp}|_{r=R_{\odot}} = {\hat{\bf r}} \times \left [ \left (
- \frac{1}{c} \, {\bf v}_0 \times
{\bf B}_{\rm torus} (R_{\odot}, \theta, \phi, t) \right )
\times {\hat{\bf r}} \right ].
\label{eq_emf}
\end{equation}
that corresponds to the upward advection at a constant velocity
${\bf v}_0$ of a magnetic field structure ${\bf B}_{\rm torus}$ given
below, defined in its own local spherical polar coordinate system
($r'$, $\theta'$, $\phi'$) whose origin is located at
${\bf r} = {\bf r}_0 = (r_0, \theta_0, \phi_0)$ of the sun's spherical
coordinate system and whose polar axis is
parallel to the polar axis of the sun's spherical coordinate system:
\begin{equation}
{\bf B}_{\rm torus} = \nabla \times \left (
\frac{A(r',\theta')}{r' \sin \theta' } \hat{\bf \phi'} \right )
+ B_{\phi'} (r', \theta') \hat{\bf \phi'},
\label{eq:Btorus}
\end{equation}
where
\begin{equation}
A(r',\theta') = \frac{1}{2} q a^2 B_t
\exp \left( - \frac{\varpi^2(r',\theta')}{a^2} \right) ,
\label{eq:afunc}
\end{equation}
\begin{equation}
B_{\phi'} (r', \theta') = \frac{a B_t}{r' \sin \theta'}
\exp \left( - \frac{\varpi^2(r',\theta')}{a^2} \right),
\label{eq:bph}
\end{equation}
where $a$ is the minor radius of the torus,
$\varpi = (r'^2 + R'^2 -2r'R' \sin \theta')^{1/2}$ is
the distance to the circular axis of the torus, in which $R'$ is
the major radius of the torus, $q/a$ denotes the rate of field
line twist (rotation in rad per unit length) about the circular
axis of the torus, and $B_t a/R'$ gives the field strength at
the circular axis of the torus.
For all the simulations in this work, we have
$a = 0.04314 R_{\odot}$ and $q/a = - 0.0166$ rad ${\rm Mm}^{-1}$.
We have used different values for the major radius $R'$ and
axial field strength $B_t a/R'$, to carry out simulations with
 different lengths and hence different total twist
of the emerged portion of the flux rope (since here we fix the
twist rate $q/a$ of the torus). The different cases of $R'$ used
will be given later in section 3.1.
Also for specifying the flux emergence
via ${\bf E}_{\perp}|_{r=R_{\odot}}$
given by (\ref{eq_emf}), it is assumed that the torus' center
position ${\bf r}_0$ is
initially located at $(r_0 = R_{\odot}-a-R', \, \theta_0 = \pi /2, \,
\phi_0 = 0) $ (thus the torus is initially entirely below the surface)
and moves bodily towards the lower boundary at a constant
velocity ${\bf v}_0 = v_0 \hat{{\bf r}}_0$, with $v_0 = 1.95$ km/s,
until a time when the emergence is stopped and
${\bf E}_{\perp}|_{r=R_{\odot}}$ is set to zero.
The velocity field at lower boundary is specified to be
uniformly $v_0 \hat{{\bf r}}_0$ in the area where the
emerging torus intersects the lower boundary and zero everywhere
else.
Note that $\hat{{\bf r}}_0$ denotes a constant unit vector
that does not change with position (unlike $\hat{\bf r}$) - it is the
direction of the position vector ${\bf r}_0$ of the center
of the torus.
The imposed advection speed $v_0$ is orders of magnitude smaller
than the Alfv\'en and the sound speed in the coronal domain to
ensure that the emerging flux rope is allowed to evolve quasi-statically
during the driving flux emergence phase.
For the side boundaries of the simulation domain, we assume
non-penetrating stress-free boundary for the velocity field
and perfectly electric conducting walls for the magnetic field.
For the top boundary, we use a simple outward extrapolating
boundary condition that allows plasma and magnetic field to flow
through.

Here we comment on the use of the semi-relativistic MHD.
The last term on the right hand side of equation \ref{eq:motion} represents
the semi-relativistic correction to the classical MHD momentum equation
when the Alfv\'en speed $v_a$ becomes relativistic (becomes comparable to the
speed of light $c$) while the bulk plasma speed and the sound speed remain
non-relativistic \citep[e.g.][]{Gombosi:etal:2002, Rempel:2017}. This is
also known as the ``Boris correction''.
It effectively increases the inertia for acceleration perpendicular to
the magnetic field and limits the Lorentz force.
In regions of the solar corona with strong magnetic field, the Alfv\'en
speed can become extremely high (approaching or even exceeding
speed of light), and therefore the semi-relativistic correction is
appropriate.  Furthermore, the extreme Alfv\'en speed can impose
stringent numerical time stepping constraint for classical MHD. 
Thus by using the semi-relativistic correction and also artificially
reducing the speed of light $c$, one can take significantly
larger time steps for numerical integration
\citep[e.g.][]{Gombosi:etal:2002, Rempel:2017}.
This is particularly useful and appropriate for the
long quasi-static evolution phase where the coronal
flux rope is built-up under the streamer as a result of flux emergence
and/or tether-cutting reconnection.
In our simulations of this work, we have set $c = 1951$ km/s, which is
always above the the peak Alfv\'en speed in the apex cross-section of the
emerged flux rope, but below the peak Alfv\'en speed for
the entire domain, which is found either in the open-field solar wind
outflow region in the lower corona or at the lower boundary,
reaching about $3000$ km/s.  
In this way we are able to take larger time steps (than that with
classic MHD) through the long quasi-static evolution phase, but still
properly model the acceleration of the flux rope during the dynamic eruption
phase.
There are two major effects of using an artificially reduced speed
of light $c$.  One is that it increases the inertia for acceleration
perpendicular to the direction of the magnetic field and thus can alter the
acceleration of the flux rope during the onset of eruption.
The other is that it also reduces the effective numerical
viscosity and diffusivity by reducing the maximum speed used for the upwinded
evaluation of the advection fluxes \citep{Rempel:2017}, and hence can
reduce the numerical diffusion during the long quasi-static phase of the
evolution.
Therefore in regard to the second effect, the simulation without the Boris
correction and the reduced speed of light may not be ``more accurate'' or
considered a ``reference'' solution.
Through test simulations with varying $c$, we find that if we reduce $c$ too
much, to below about 1/2 of the Alfv\'en speed
of the flux rope apex cross section, we begin to see significant decrease of
the peak acceleration of the flux rope during the onset of the eruption.
We find that if we use $c$ comparable to or greater than the peak Alfv\'en
speed in the apex cross-section of the flux rope
(as is the case here with $c = 1951$ km/s),
the dynamic evolution of the eruption
remains very close to that obtained without the Boris correction.

\clearpage
\section{Results}
\subsection{The initial helmet streamer fields}
\begin{figure}[htb!]
\plottwo{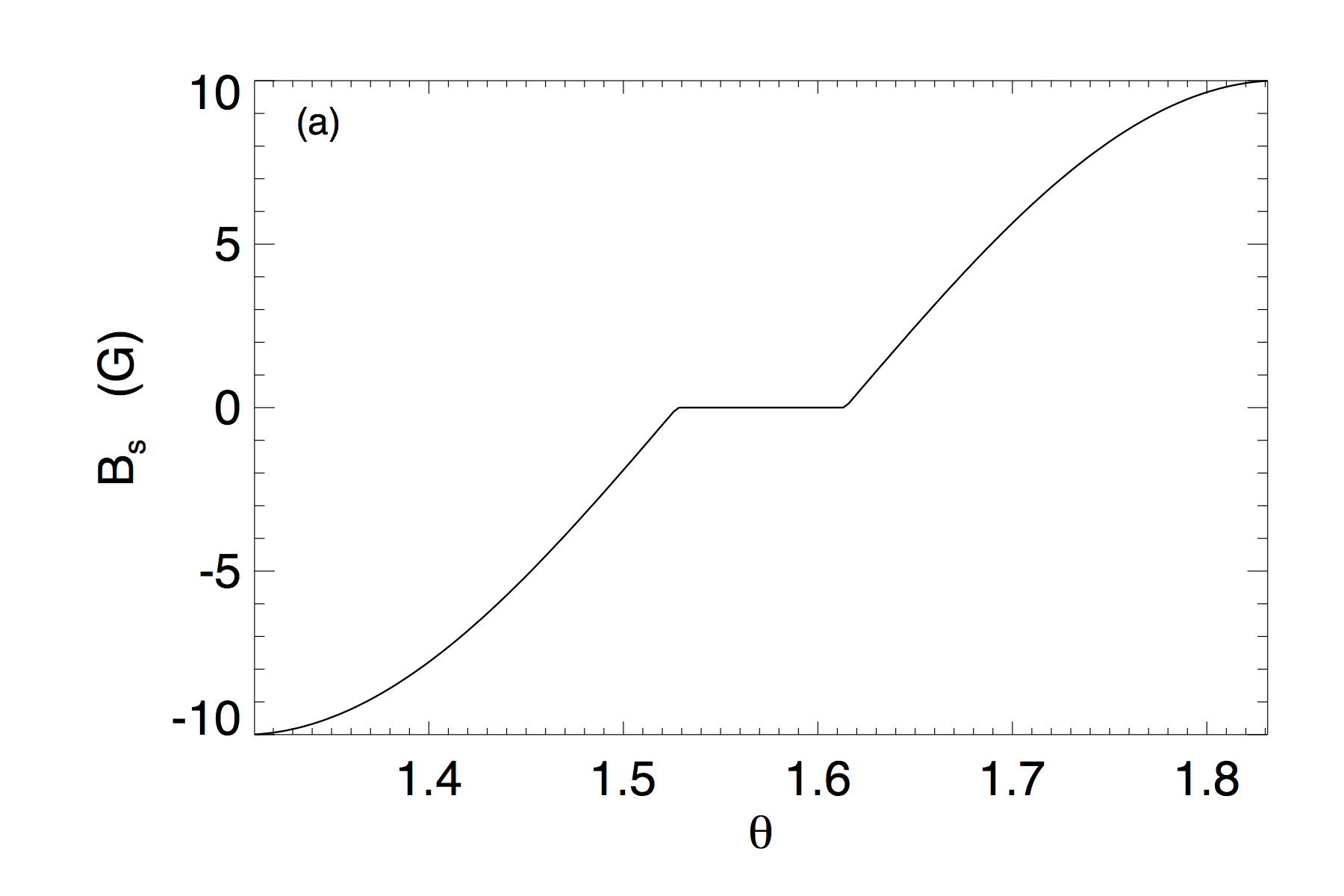}{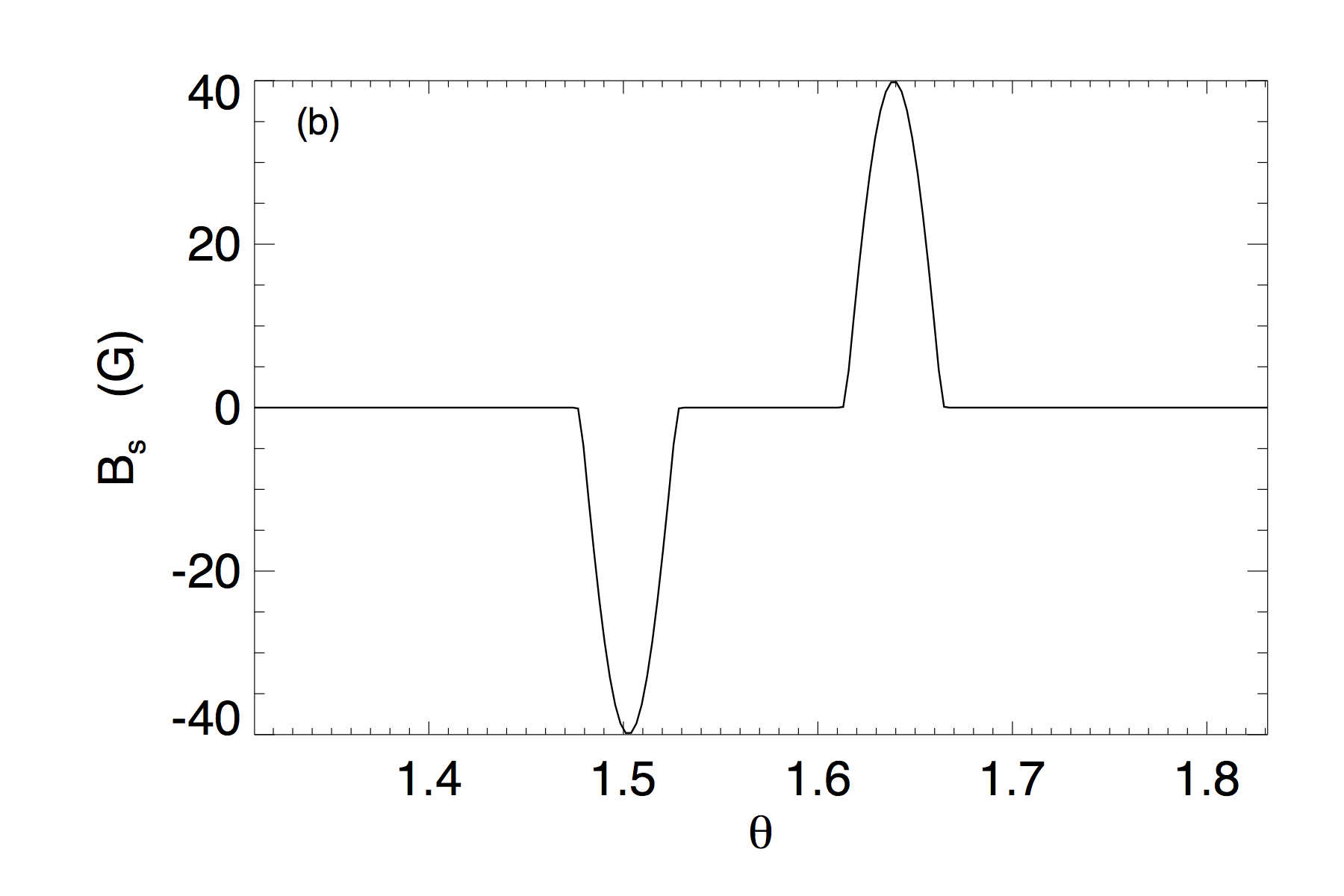}
\caption{The lower boundary normal magnetic field
$B_s (\theta)$ used for the WS (panel a) and NS (panel b) initial
streamer solutions}
\label{fig:initbr}
\end{figure}

\begin{figure}[htb!]
\centering
\includegraphics[width=0.3\textwidth]{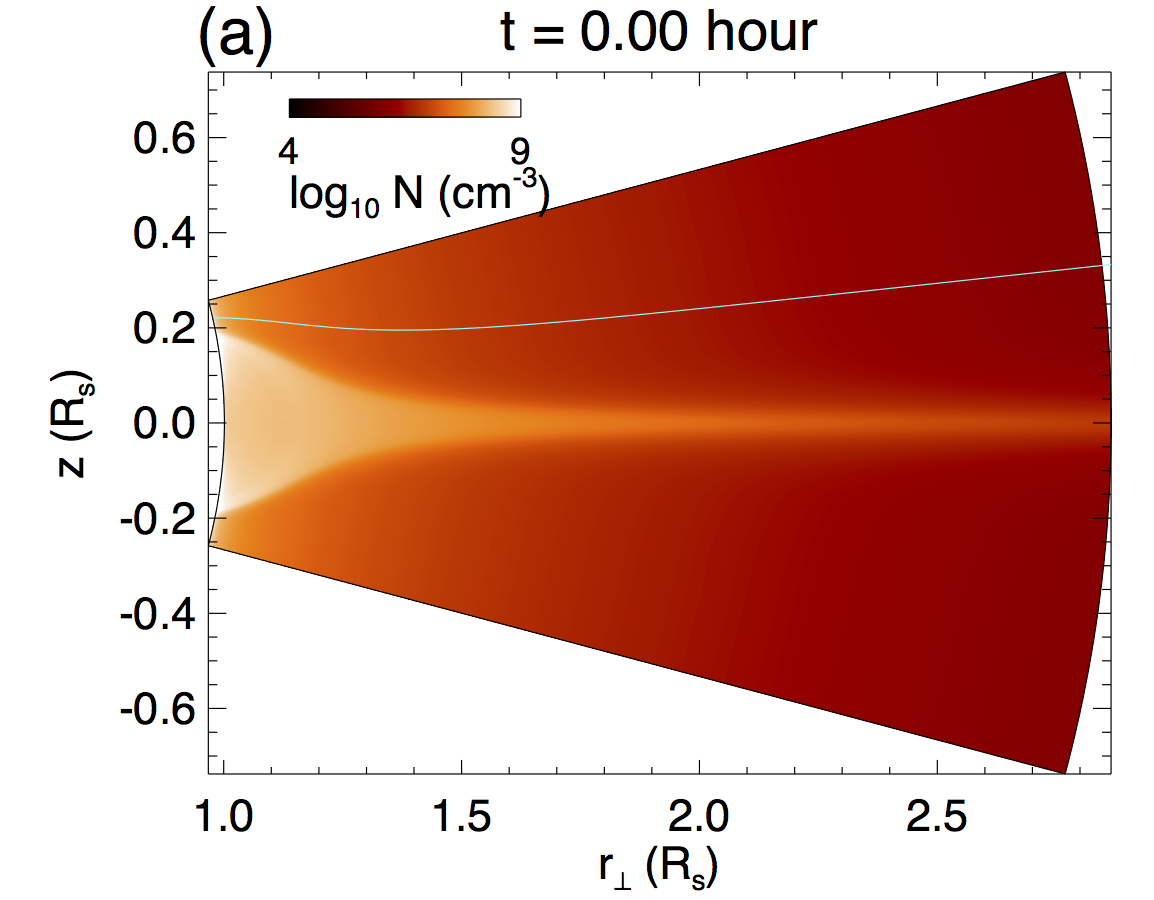}
\includegraphics[width=0.3\textwidth]{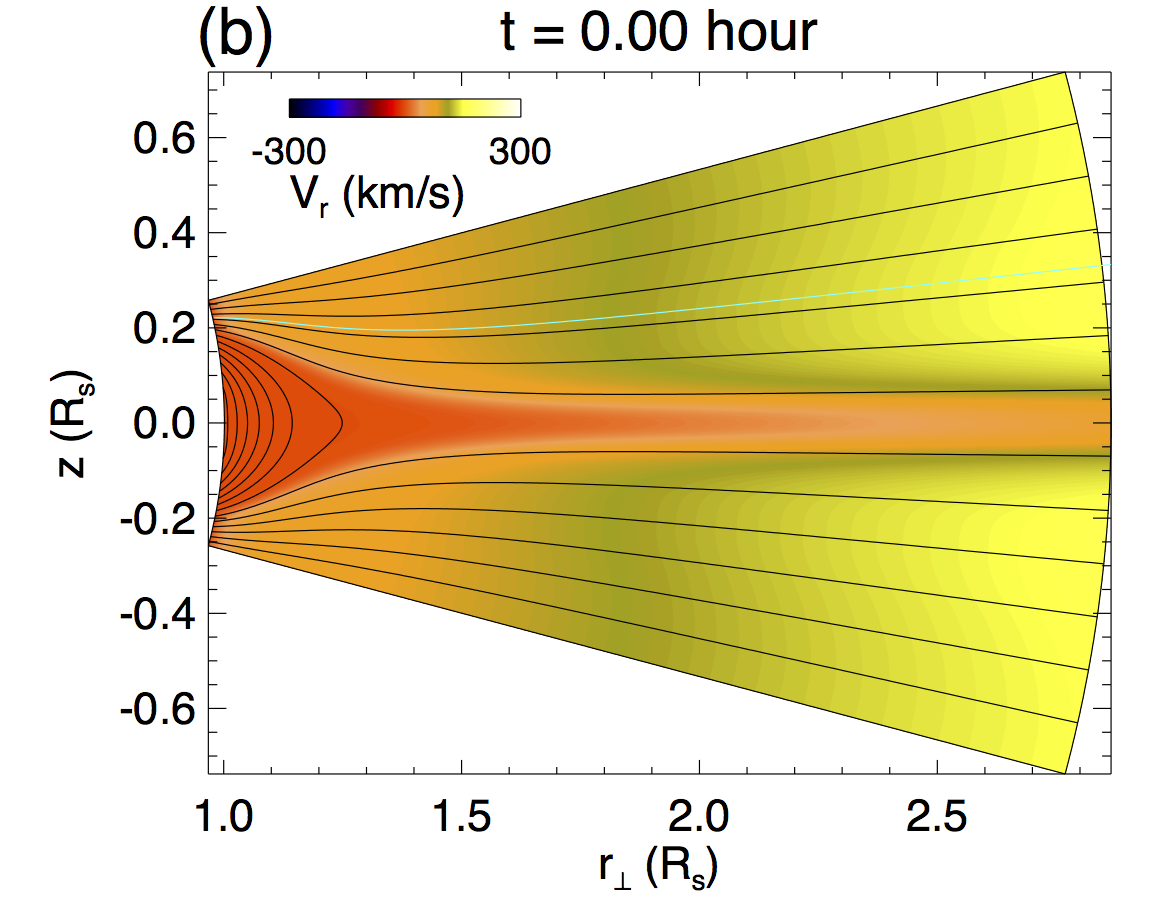} \\
\includegraphics[width=0.3\textwidth]{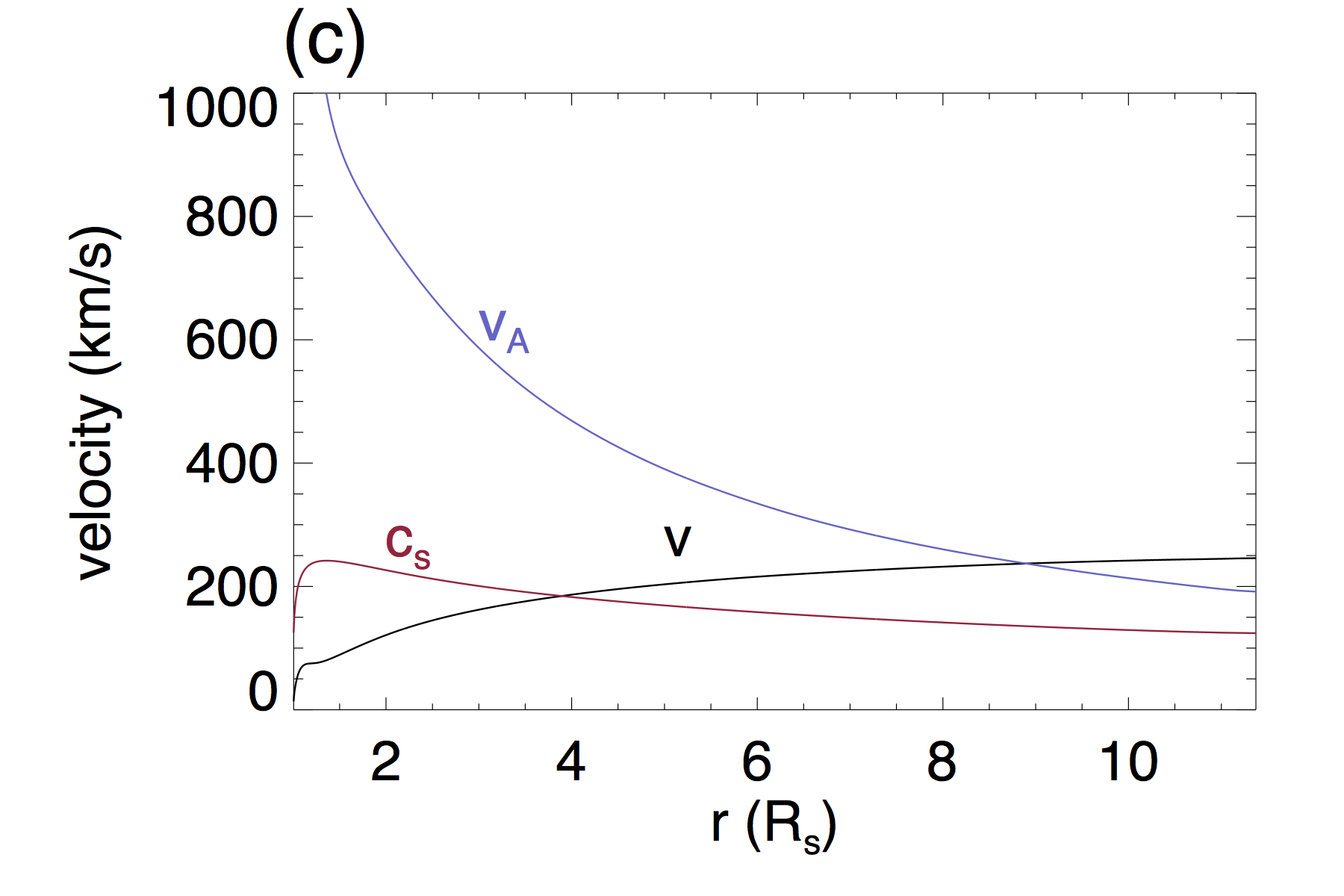}
\includegraphics[width=0.3\textwidth]{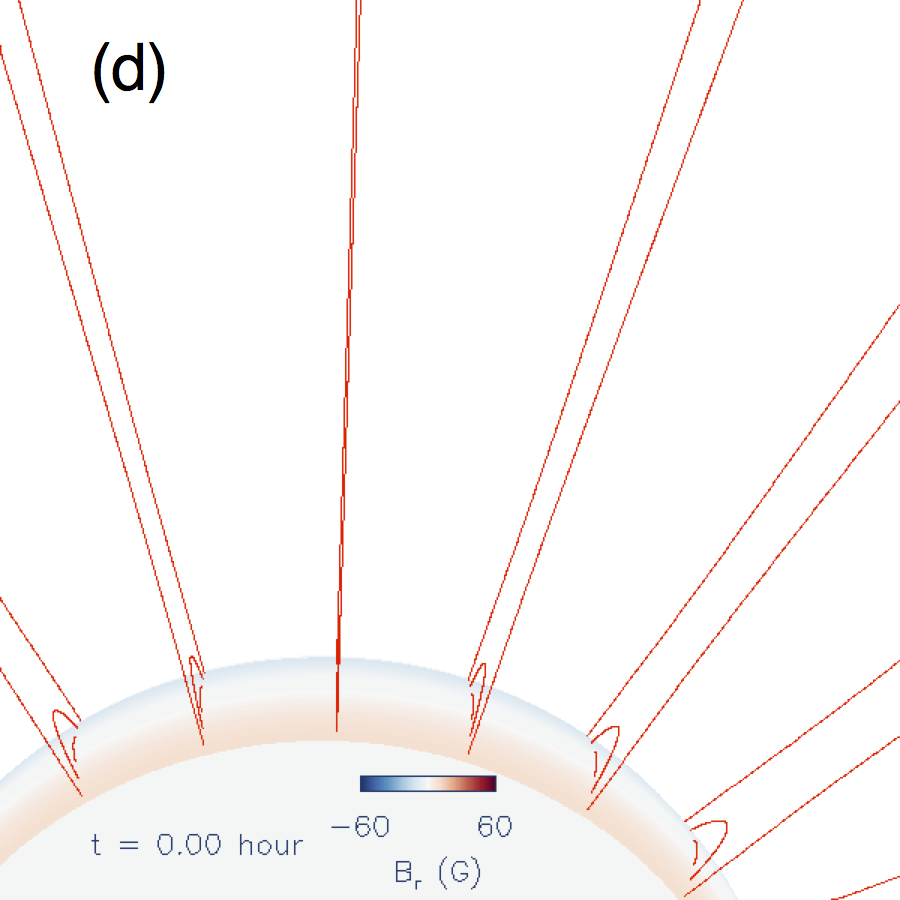}
\caption{The WS initial state: (a) shows the density in the meridional cross-section;
(b) shows the radial velocity in the meridional cross-section over plotted with
contours of the magnetic field lines; (c) shows the parallel velocity $V$,
the Alfv\'en speed $V_a$, and the sound speed $C_s$, along an open field line
(the green line shown in panel (a) and (b));
(d) shows a 3D view of the initial field lines in the simulation domain
with the lower boundary spherical surface colored based on the normal
magnetic field $B_r$.}
\label{fig:WS}
\end{figure}

For the initial state of the simulations, we initialize two different
2D quasi-steady solutions of a coronal streamer with a background solar
wind. We initialize a wide streamer (WS) solution and a
narrow streamer (NS) solution, for which the normal magnetic flux
distributions at the lower boundary are bipolar bands with
$B_r(R_{\odot}, \theta, \phi) = B_s (\theta)$, where $B_s (\theta)$ used for
the WS and NS solutions are shown in Figure \ref{fig:initbr}(a) and
Figure \ref{fig:initbr}(b) respectively.
With the lower boundary normal magnetic field distribution given above, we
first construct an initial potential magnetic field together with a
hydrostatic atmosphere with a specified temperature profile that increases
linearly from $5 \times 10^5$ K at the lower boundary to $1.5 \times 10^6$
K at $22.5$ Mm height, and then remains constant to the outer 
boundary at about $11.5 R_{\odot}$.
The initial potential field is a 2D arcade field (invariant in $\phi$)
given by ${\bf B}_p (r, \theta) = \nabla \Phi (r, \theta)$,
where the potential $\Phi$ satisfies the Laplace equation
$\nabla^2 \Phi (r, \theta ) = 0$. Discretizing the
2D Laplace equation for $\Phi$ with the appropriate boundary conditions
lead to a block tridiagonal system, which is solved using the routine
blktri.f in the FISHPACK math library of the National Center for
Atmosphereic Research (NCAR) based on the generalized cyclic reduction
scheme developed by P. Swatztrauber of NCAR.
We then lower the pressure at the outer boundary of the initial static
state to initiate an outflow, and let the
system relax with time (following the MHD equations) until a quasi-steady
state is reached and the potential field is stretched out into a
streamer configuration.
The relaxed WS initial streamer solution is shown in Figure \ref{fig:WS},
\begin{figure}[hb!]
\centering
\includegraphics[width=0.3\textwidth]{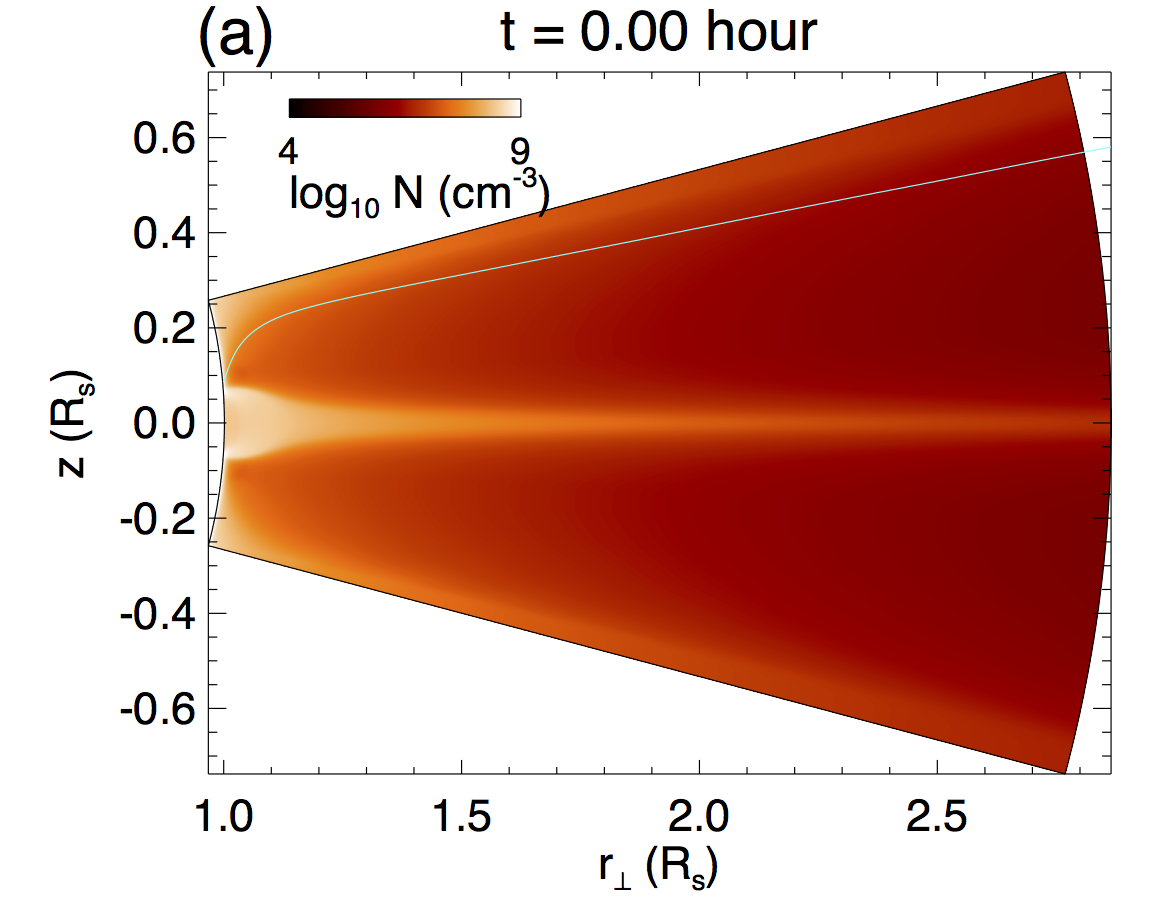}
\includegraphics[width=0.3\textwidth]{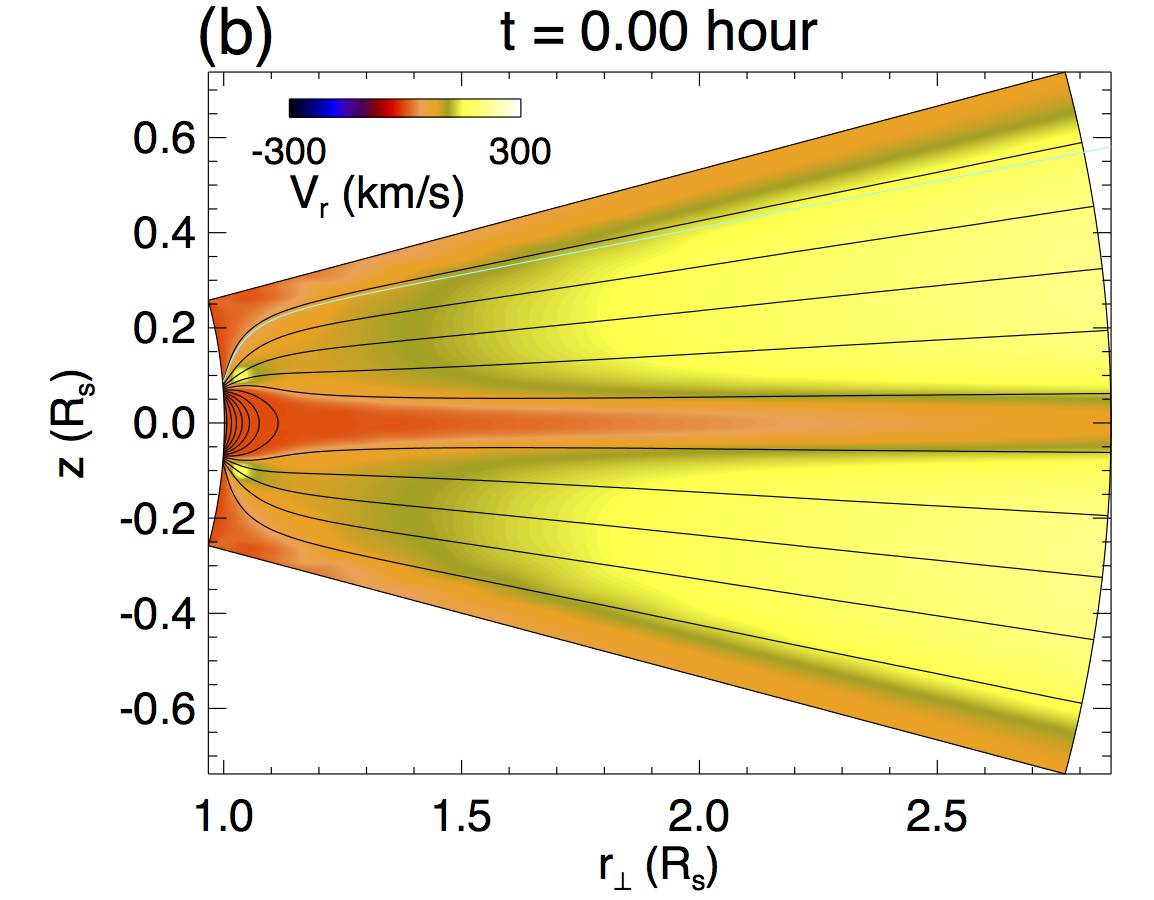} \\
\includegraphics[width=0.3\textwidth]{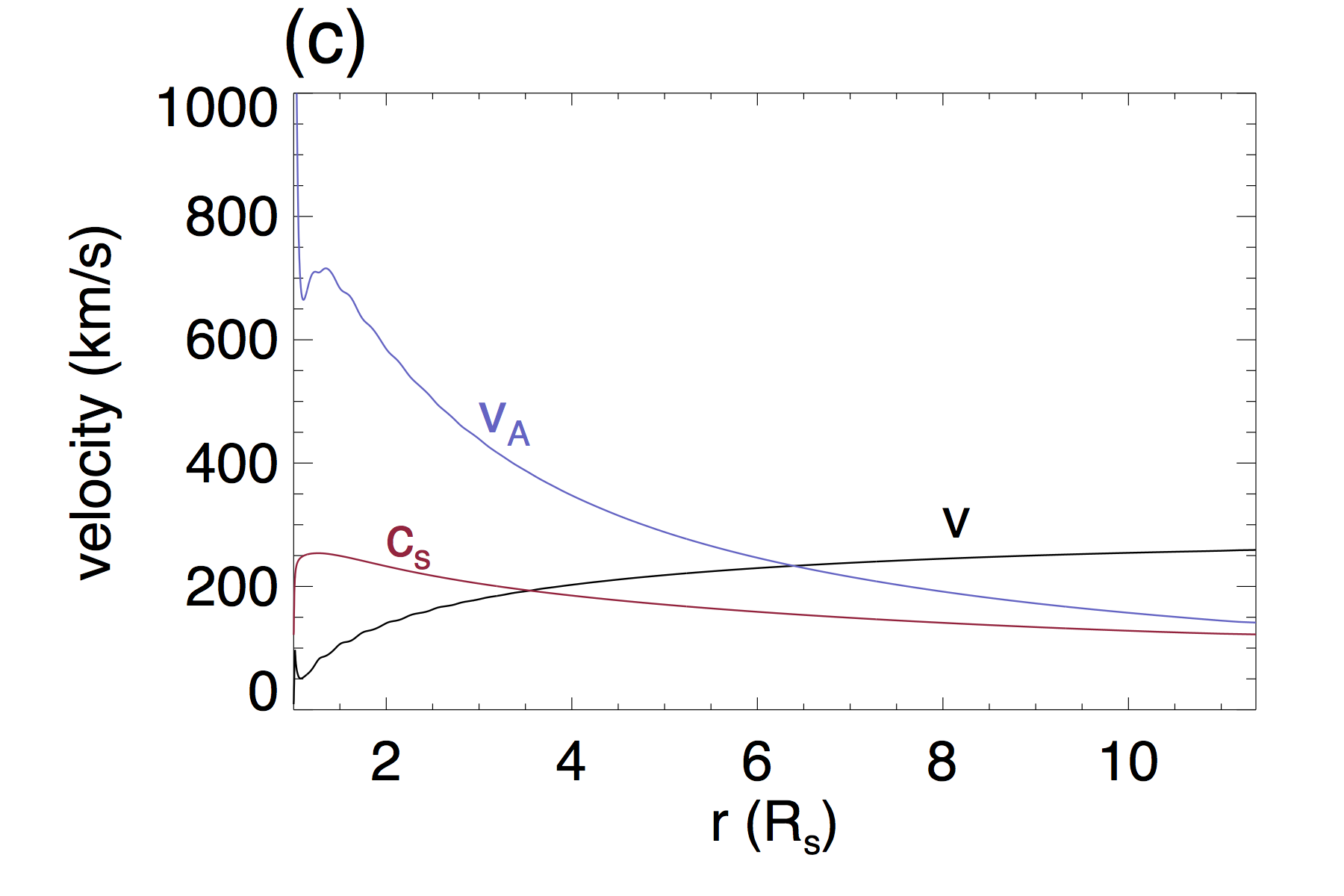}
\includegraphics[width=0.3\textwidth]{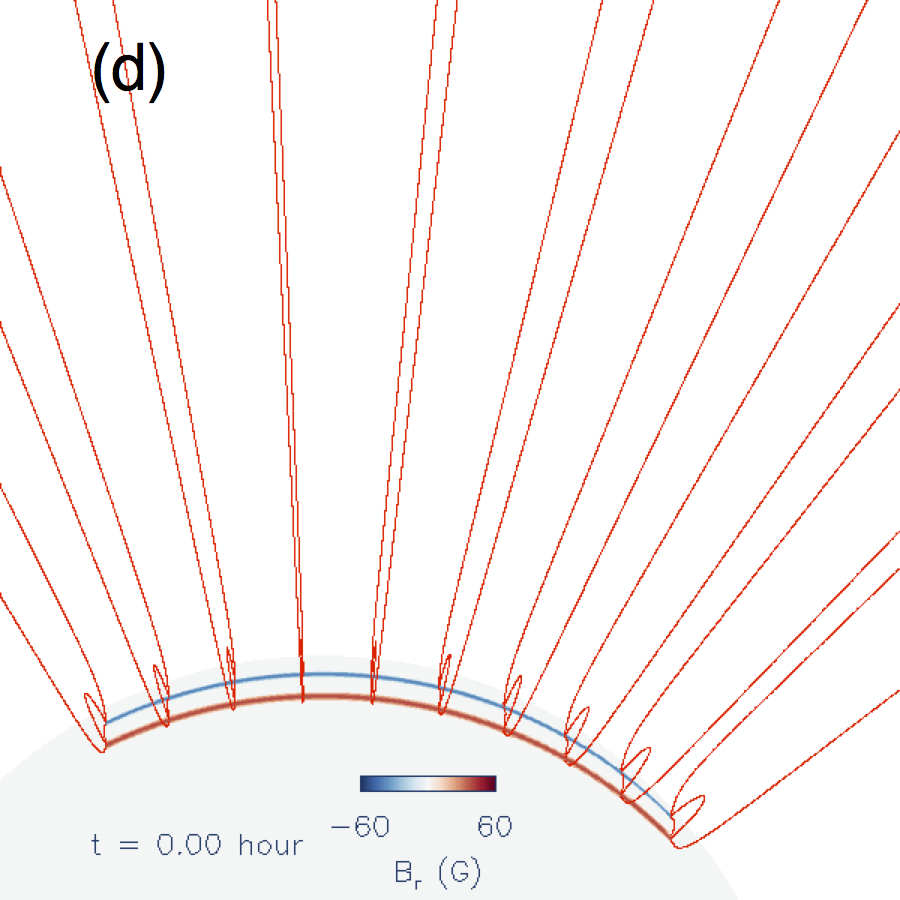}
\caption{Same as Fig. \ref{fig:WS} but for the NS initial state.}
\label{fig:NS}
\end{figure}
where we see a helmet streamer with a denser helmet
dome of closed magnetic field in approximately static equilibrium, in
an ambient low density open field region with a solar wind outflow.
The outflow speed, and the sound speed and the Alfv\'en speed
 along an open field line (indicated by the green field line in
the top two panels) are shown in the bottom panel of Figure
\ref{fig:WS}. The wind speed becomes super-sonic at about
$4 R_{\odot}$ and super-Alfv\'enic at about $9 R_{\odot}$,
reaching a peak speed of about 250 km/s at the outer boundary
at about $11.5 R_{\odot}$.
The solar wind obtained here is a thermally driven wind heated
by the highly simplified empirical coronal heating function given
in equation \ref{eq:heatingfunc}. The peak wind speed reached at
$11.5 R_{\odot}$ is significantly lower than that reached by
the fast wind in solar coronal holes because here we have not
included the acceleration and heating by the Alfv\'en wave
turbulence \citep[e.g.][]{vanderHolst:etal:2014}.
However, with our simple initial solar wind solution,
we obtain the partially open coronal magnetic field
configuration of a helmet streamer with a denser helmet dome
compared to the open field region, in qualitative
agreement with observations.
Similarly the NS initial streamer solution is shown in
Figure \ref{fig:NS}, where we obtained a significantly smaller
and narrower streamer configuration by using a lower boundary
normal flux distribution (Figure \ref{fig:initbr}(b)) of a
narrower and thinner pair of bipolar bands.
The solar wind speed in the open field region is similar to
that for the wider steamer case.

Into the dome of the initial streamer field configuration, we
drive the emergence of an arched flux rope with a varying length
and (hence) total twist, to study the subsequent evolution of
the transition from quasi-equilibrium to eruption of the helmet
dome.
We vary the length and total twist
of the emerged flux rope by changing the curvature radius $R'$
of the magnetic torus ${\bf B}_{\rm torus}$ used in specifying the
the lower boundary electric field for driving the flux emergence
as described in section 2.
The three numerical simulations carried out, where we use
either the WS and NS solution as the initial state, and impose the
emergence of the magnetic torus with the different curvature radius
$R'$ to obtain an emerged flux rope with different length and total
twist, are summarized in Table \ref{table:cases}.
The label of each run is such that the first 2 letters are
either ``WS'' or ``NS'' denoting which initial helmet streamer
solution is used, and the 3rd letter is ``L'' (for Long),
``M'' (for Medium), or ``S'' (for Short), which correspond to
setting the curvature radius $R'$ to $0.75 R_{\odot}$,
$0.5 R_{\odot}$, or $0.25 R_{\odot}$ respectively.
The other varying parameters used for each of the runs, i.e. 
the axial field strength $B_t a/R'$
of the emerging torus used for specifying the lower boundary
electric field, the domain size $\phi_{\rm max}$ in the
azimuthal(longitudinal) direction, and the total field line twist reached
in the emerged flux rope when the emergence is stopped, are also listed
in Table \ref{table:cases}.
\begin{deluxetable}{cccccc}[htb!]
\tabletypesize{\scriptsize}
\tablecaption{Summary of simulations\label{table:cases}}
\tablewidth{0pt}
\tablehead{\colhead{case label} & \colhead{initial streamer} &
\colhead{$R'$ ($R_{\odot}$) \tablenotemark{a}} &
\colhead{$B_t a/R' $ (G) \tablenotemark{b}} &
\colhead{$\phi_{\rm max}$ \tablenotemark{c}} &
\colhead{emerged twist (winds) \tablenotemark{d}}}
\startdata
WS-L & wide streamer & 0.75 & 100 & $75^{\circ}$ & 1.83 \\ 
WS-M & wide streamer & 0.5 & 103 & $37.5^{\circ}$ & 1.1 \\ 
NS-S & narrow streamer & 0.25 & 90 & $37.5^{\circ}$ & 0.6 \\
\enddata
\tablenotetext{a}{curvature radius of the torus}
\tablenotetext{b}{axial field strength of the torus}
\tablenotetext{c}{domain size in $\phi$: $[-\phi_{\rm max},\phi_{\rm max}]$}
\tablenotetext{d}{total field line twist about the axial field line
in the corona between the anchored ends when the emergence is stopped}
\end{deluxetable}

\subsection{Eruption under a wide and a narrow streamer}
\begin{figure}[htb!]
\centering
\includegraphics[width=0.85\textwidth]{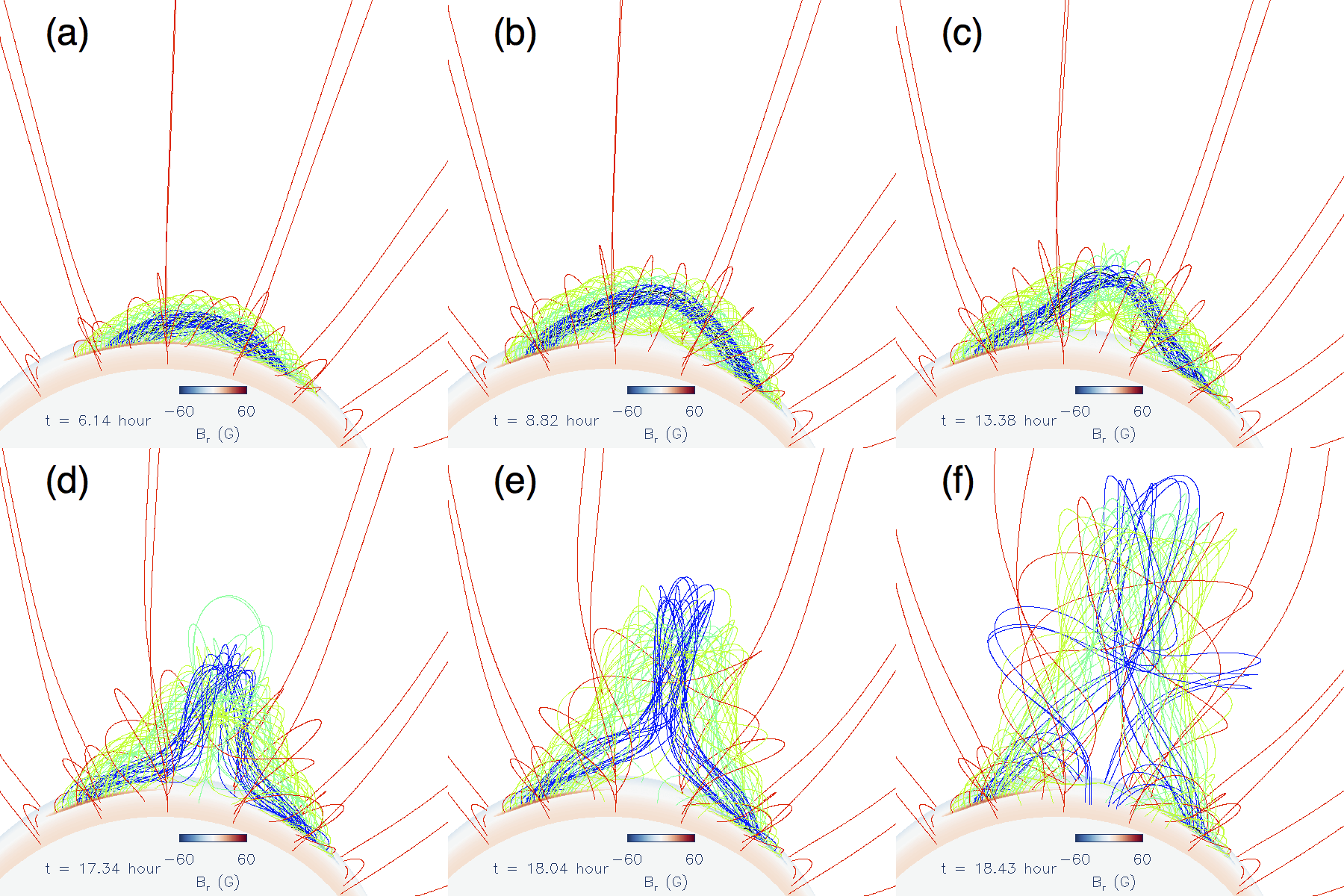} \\
\includegraphics[width=0.35\textwidth]{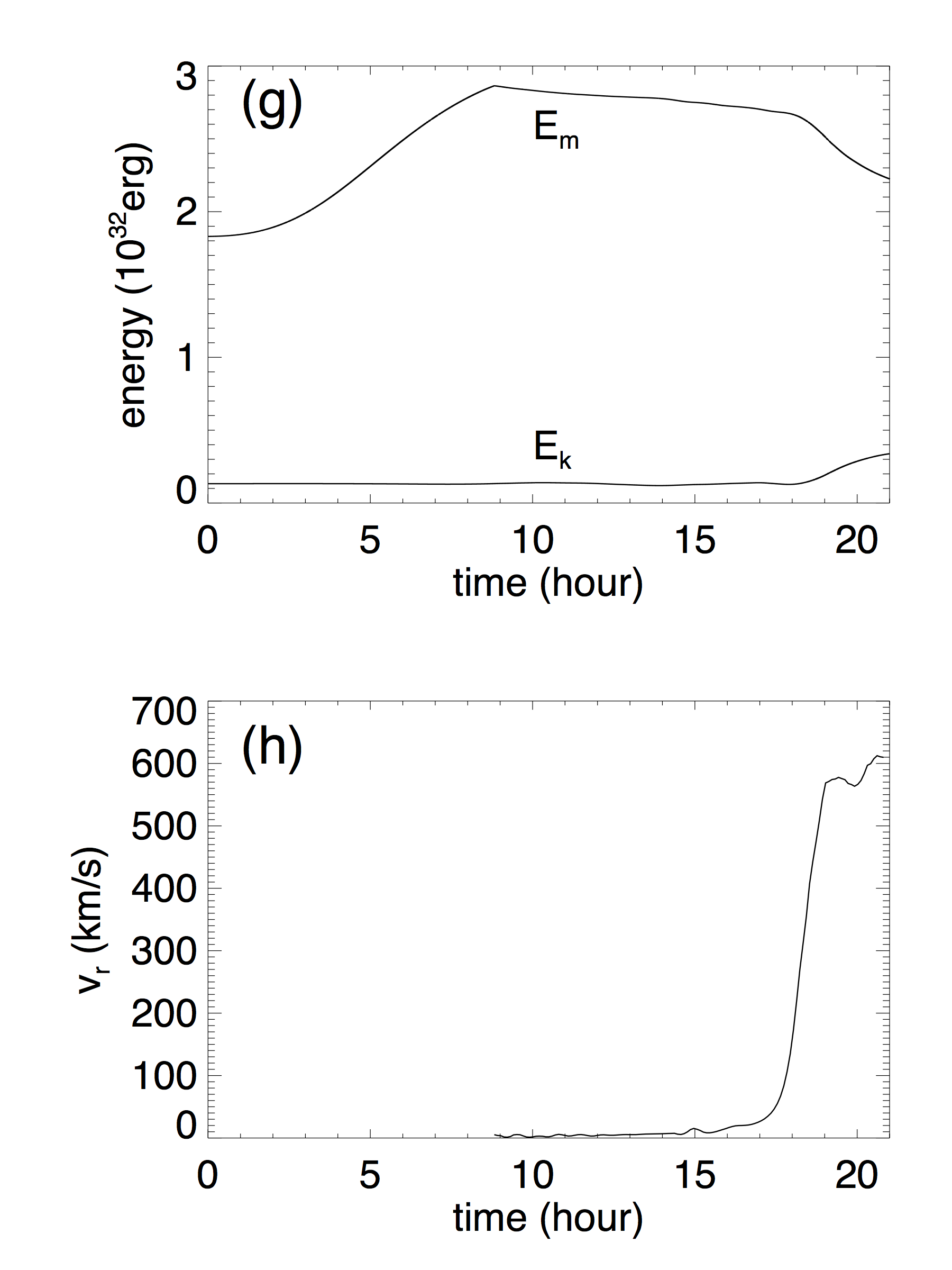}
\caption{The evolution obtained from simulation case WS-L. Panels (a)-(f)
show snapshots of the 3D magnetic field lines. The lower boundary surface
is colored with $B_r$. The color of the field lines
are based on the original flux surfaces as described in the text.
Panel (g) shows the evolution of the total magnetic energy $E_m$ and total kinetic energy
$E_k$. Panel (h) shows the evolution of the rise velocity $v_r$ tracked at the
apex of the emerged flux rope axis. A movie for this figure
showing the evolution of the 3D magnetic field, and the evolution of $E_m$, $E_k$, and $v_r$
throughout the course of the simulation (from $t=0$ to about $t=21$ hour)
is available in the online version of the paper.}
\label{fig:fdl3d_WS-L}
\end{figure}
Panels (a)-(f) of Figure \ref{fig:fdl3d_WS-L} show the 3D magnetic field evolution obtained
from the simulation case WS-L, where we drive the emergence of a long flux
rope at the lower boundary into the wide streamer initial state in an azimuthally
long domain.
Panels (g) and (h) of Figure \ref{fig:fdl3d_WS-L} show the corresponding evolution of the
magnetic energy $E_m$, the kinetic energy $E_k$, and the rise velocity $v_r$ tracked at the apex
of the axial field line of the emerged flux rope.
A movie corresponding to Figure \ref{fig:fdl3d_WS-L} showing the 3D field evolution and
the evolution of $E_m$, $E_k$, and $v_r$, is available in the online version
of the paper.
The field lines and their colors are selected in the following way.
We use a set of fixed foot points in the pre-existing arcade bands outside the
emerging flux region and trace the field lines in red. For tracing the field lines
(green, blue, and black field lines) from the emerging flux region on the lower
boundary, we track a set of foot points that connect to a fixed set of field
lines of the subsurface emerging torus and color the field lines based on the
flux surfaces of the subsurface torus.
The ``axial field line'' refers to the field line that is traced from the
footpoints at the lower boundary that connect to the curved axis of the subsurface
emerging torus ${\bf B}_{\rm torus}$.
We track the apex position of this field line for evaluating the $v_r$ in panel (h) of Figure
\ref{fig:fdl3d_WS-L}.
After the axial field line has reconnected during its coronal evolution,
we continue to track the Lagrangian evolution of this apex element by using
its velocity, and continue to refer to it as the ``apex of the axial field line''.
In panel (g) of Figure \ref{fig:fdl3d_WS-L}, we see initially
(from about $t=0$ hour to $t=8.82$ hour) the magnetic energy
increases as the coronal flux rope is built up
quasi-statically under the streamer
(see also panels (a) and (b) of Figure \ref{fig:fdl3d_WS-L})
as a result of flux emergence driven at the lower boundary.
The driving flux emergence is stopped at $t=8.82$ hour when the total field
line twist about the axial field line of the emerged flux rope reaches
1.83 winds, which is above the critical twist (about $1.25$ winds)
for the onset for the kink instability \citep{Hood:Priest:1981}.
Subsequently the flux rope becomes kinked due to the development of the helical
kink instability (panels (c) and (d) of Figure \ref{fig:fdl3d_WS-L}).
However the rope remains confined with its apex rising
quasi-statically at a low, significantly sub-Alfv\'en speed
and $E_m$ decreases slowly,
until roughly $t= 17.5$ hour (see panels (g) and (h) of Figure \ref{fig:fdl3d_WS-L}),
when the flux rope's rise speed begins a significant acceleration,
$E_m$ begins a rapid decrease and $E_k$ begins a significant increase.
The period of slow rise phase (from $t=8.82$ hour to roughly $t=17.5$ hour)
after the emergence is stopped is
found to be significantly longer than the
Alfv\'en transit time $\tau_A \approx 0.136$ hour along the flux rope
(estimated by computing the Alfv\'en transit time along the axial field
line between the anchored foot points at the time the emergence is stopped),
which is a measure of the dynamic time scale.
The slow rise phase lasts about 64 $\tau_A$.
This suggests that the flux rope remains in quasi-equilibrium for the
slow rise phase. There is a clear transition
at the time of roughly $t=17.5$ hour from the slow rise phase to a
dynamic eruption phase, as can be seen in panels (g) and (h) in Figure \ref{fig:fdl3d_WS-L}.
The flux rope accelerates to a terminal speed of about $600$ km/s,
much higher than the ambient solar wind speed,
and begins to exit the domain upper boundary at $11.47 R_{\odot}$ at
about $t=21$ hour.

\begin{figure}[ht!]
\centering
\includegraphics[width=0.85\textwidth]{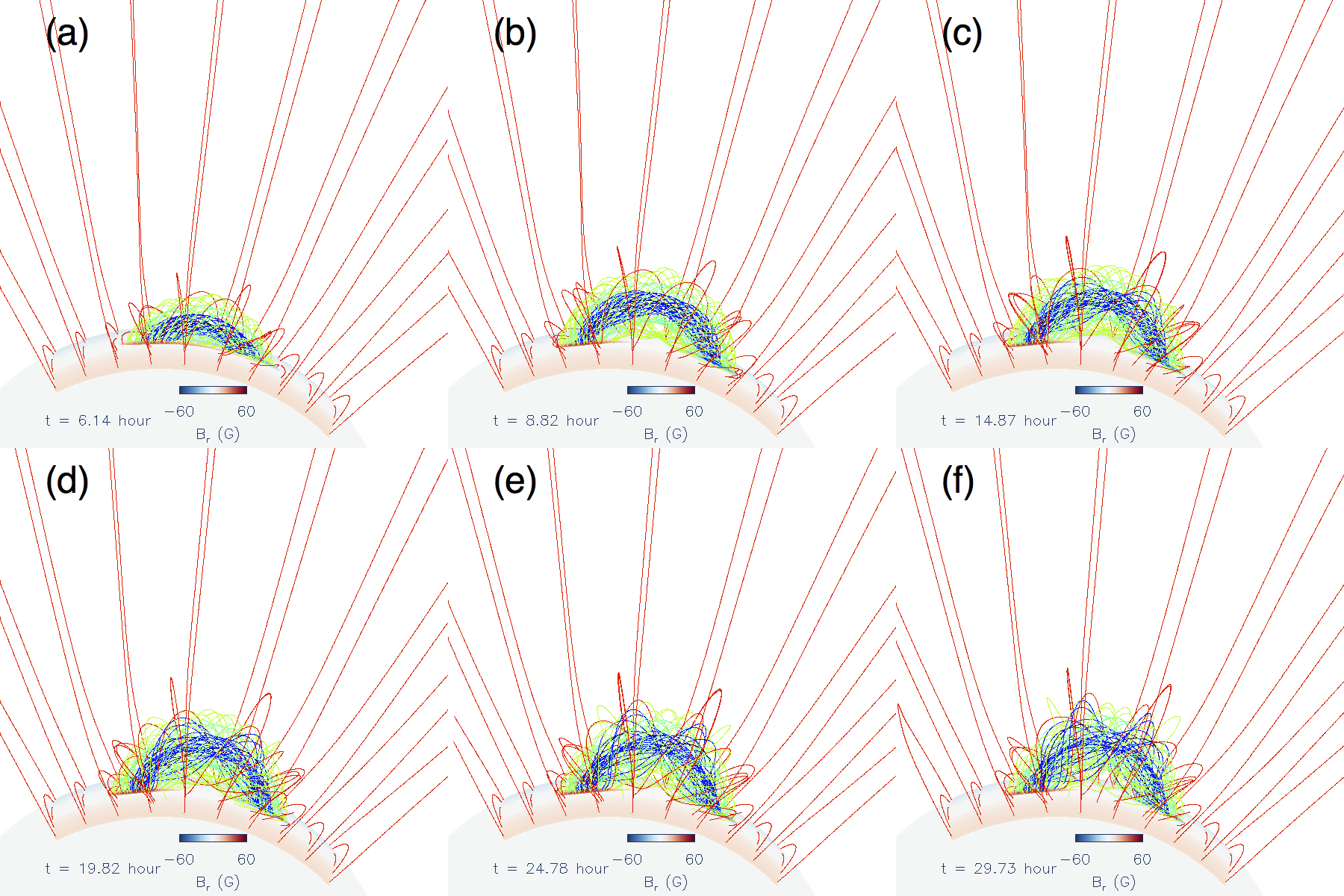}
\includegraphics[width=0.35\textwidth]{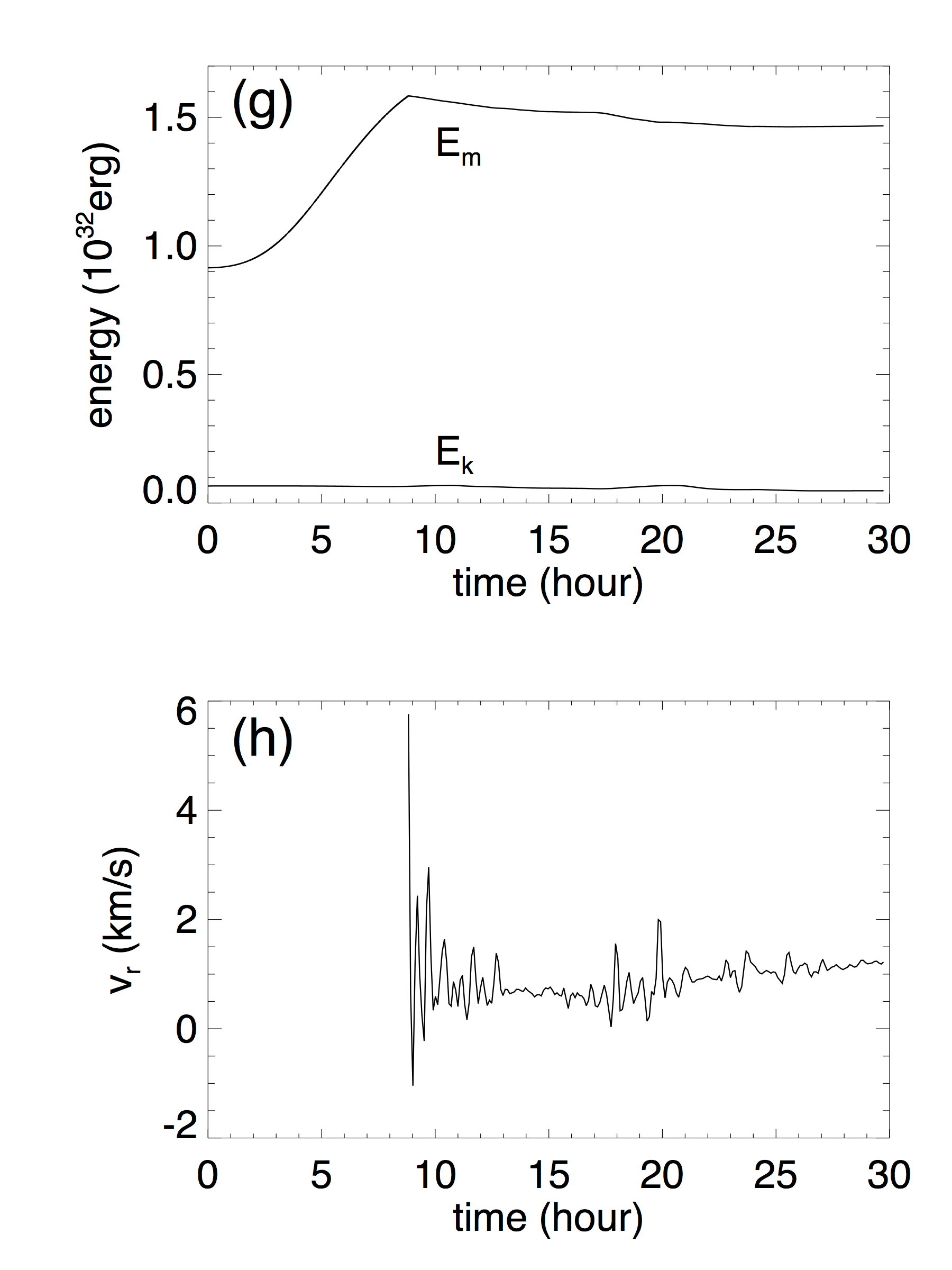}
\caption{Same as Figure \ref{fig:fdl3d_WS-L} but for the simulation case
WS-M.}
\label{fig:fdl3d_WS-M}
\end{figure}

The 3D coronal magnetic field evolution of another simulation
case WS-M is shown in Figure \ref{fig:fdl3d_WS-M},
for which we drive the emergence of a shorter, anchored flux rope,
and stop the emergence when the total field line twist about
the axial field line reaches 1.1 winds between the anchored ends (at $t=8.82$ hour,
Figure \ref{fig:fdl3d_WS-M}(b)).
For this case we see that the flux rope remains
stable and well confined in quasi-equilibrium under the streamer for the subsequent
evolution of over 20 hours simulated, which corresponds to about $268 \tau_A$, where
$\tau_A$ is the Alfv\'en transit time along the flux rope estimated at the time
the emergence is stopped, showing no sign of eruption (see panels (c)-(f) in Figure
\ref{fig:fdl3d_WS-M}). Panels (g) and (h) of Figure \ref{fig:fdl3d_WS-M}
show the evolution of the
magnetic energy $E_m$, the kinetic energy $E_k$, and the rise velocity $v_r$ tracked
at the apex of the flux rope axial field line. Again we see the build up of the
magnetic energy in the flux emergence phase (from $t=0$ to $t=8.82$ hour),
when the emergence of the magnetic torus is imposed at the lower boundary.
However, subsequently after the emergence is stopped, the rise velocity remains
very small, showing some small oscillations (panel (h) of
Figure \ref{fig:fdl3d_WS-M}).
The magnetic energy $E_m$ shows a slow decline which becomes steady later,
and the total kinetic energy $E_k$ remains fairly steady (with the ambient solar wind).
All these indicate that the magnetic flux rope in this case is settling into
a stable equilibrium under the helmet.
The results of the above two simulation cases indicate that the wide-streamer (WS)
pre-existing field is very confining, such that the flux rope does not erupt
until sufficiently high twist (significantly higher than 1 wind of field line
twist) is built up for the kink instability to set in,
which brings the apex of the kinked flux rope to a rather high height for it
to erupt dynamically.

\begin{figure}[ht!]
\centering
\includegraphics[width=0.85\textwidth]{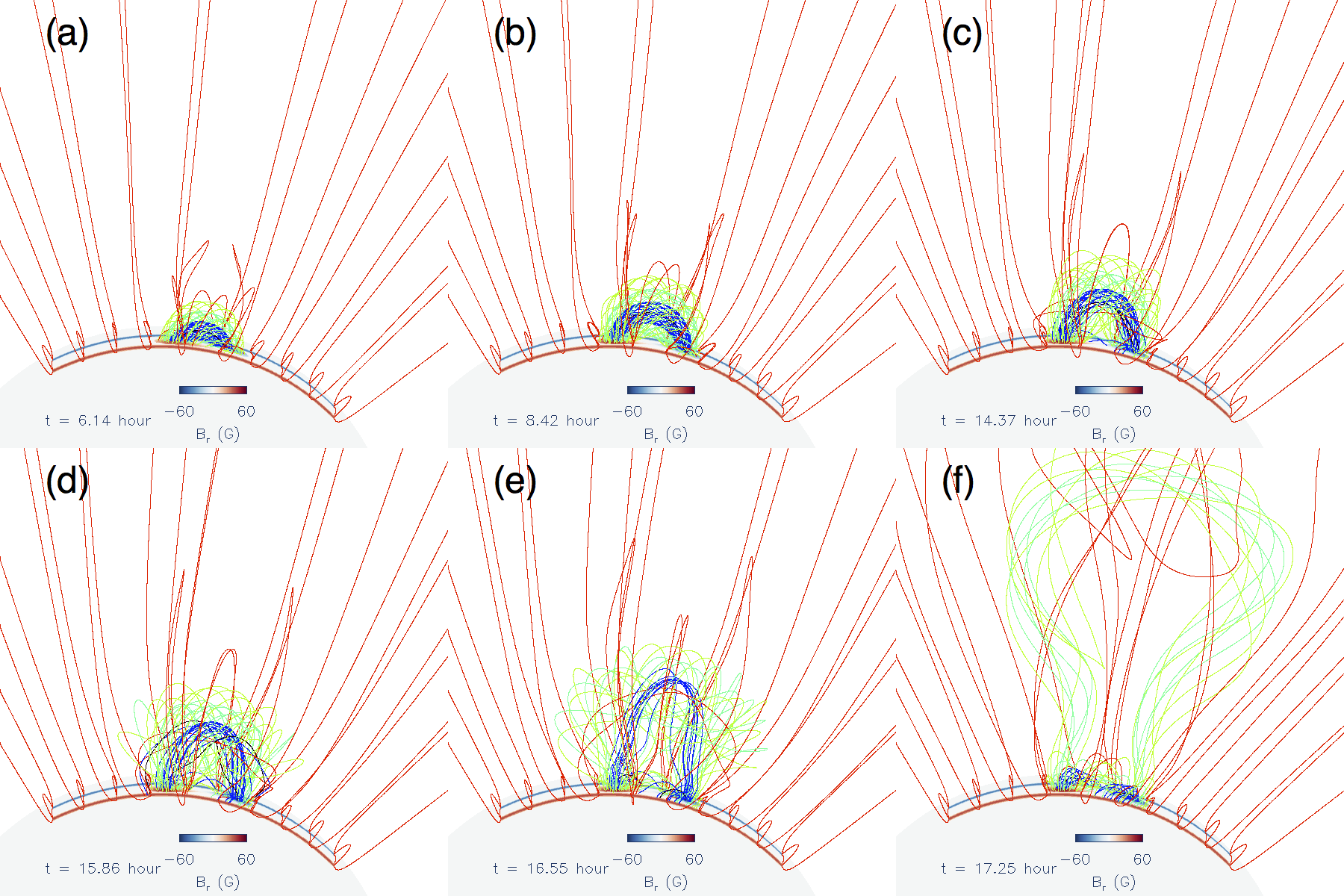}
\includegraphics[width=0.35\textwidth]{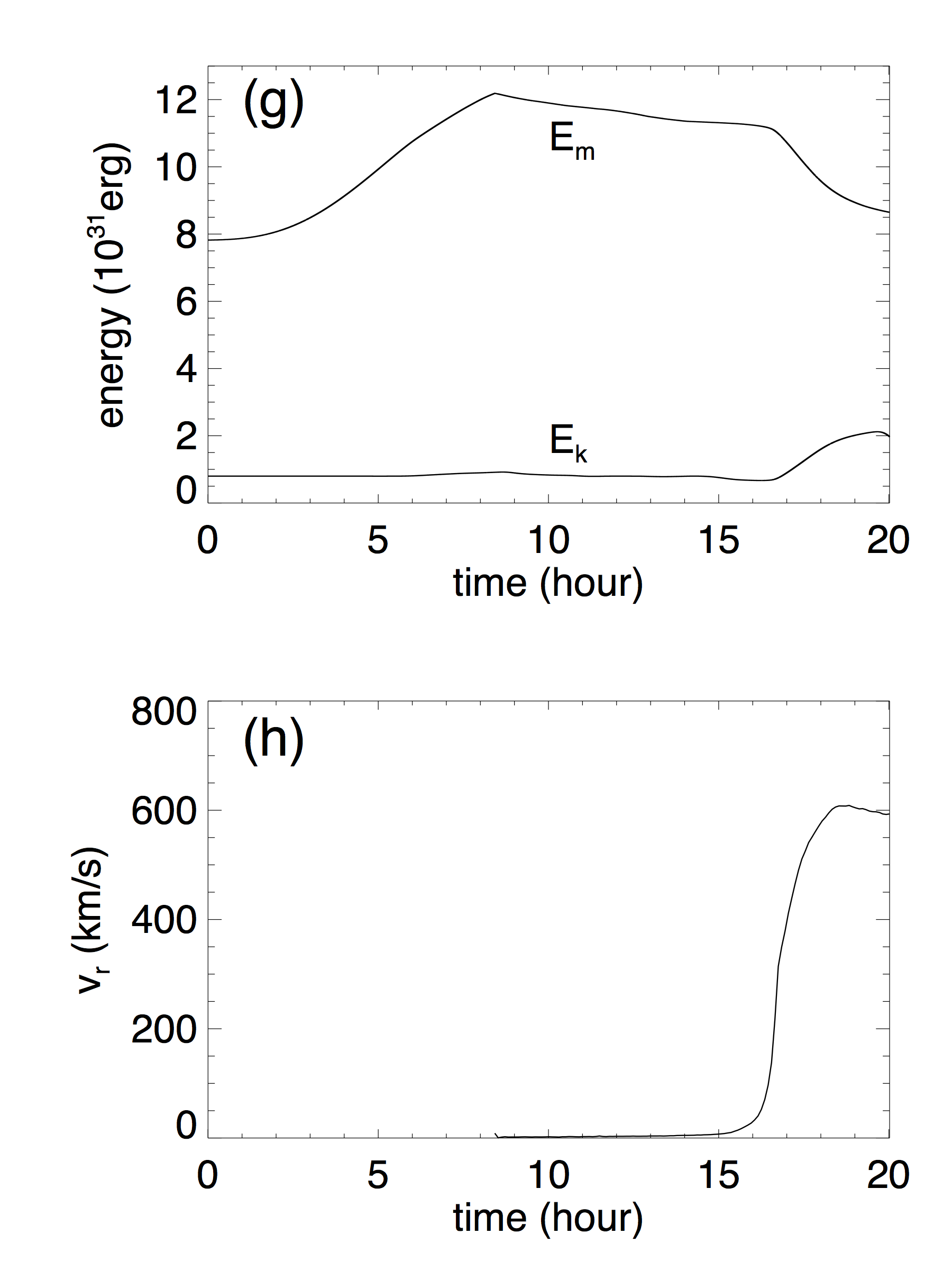}
\caption{Same as Figure \ref{fig:fdl3d_WS-L} but for the simulation case
NS-S. A movie for this Figure is
available in the online version of the paper.}
\label{fig:fdl3d_NS-S}
\end{figure}

In contrast, Figure \ref{fig:fdl3d_NS-S} shows the evolution for the simulation
case NS-S, where we use the narrow streamer field (NS) as the initial state,
and drive the emergence of an even shorter flux rope with the total field line
twist about the axial field line reaching only about 0.6 winds between the
anchored ends when the emergence is stopped (panel (b) of Figure \ref{fig:fdl3d_NS-S}).
Panel (g) and (h) of Figure \ref{fig:fdl3d_NS-S} again show the evolution of the
magnetic energy $E_m$, the kinetic energy $E_k$, and the rise velocity $v_r$ tracking
the Lagrangian element at the apex of the flux rope axial field line.
Even though in this case the twist is well below that for the onset of the kink
instability when the emergence is stopped, we see subsequently, similar to
the WS-L case, the flux rope undergoes a stage of slow, quasi-static rise with
sub-Alfv\'enic speed for a time span (from about $8.82$ hour to about $16$
hour in panel (h) of Figure \ref{fig:fdl3d_NS-S},
also panels (b)-(d) in Figure \ref{fig:fdl3d_NS-S})
that is significantly longer than the Alfv\'en crossing time $\tau_A$,
about $180 \tau_A$.  At roughly $t=16$ hour, a transition to dynamic eruption occurs
with the flux rope undergoes a significant acceleration to a terminal speed of
about $600$ km/s and with the total magnetic energy $E_m$ showing a significant
decrease and the total kinetic energy $E_k$ showing a significant increase (see
panels (g) and (h) in Figure \ref{fig:fdl3d_NS-S}
and panels (d)-(f) of Figure \ref{fig:fdl3d_NS-S}).
Since the total twist of the flux rope is significantly below the critical limit
for the onset of the kink instability, the dynamic eruption in this case is
most likely due to the onset of the torus instability when the flux rope rises to
a critical height of sufficiently steep decline of the corresponding potential
field \citep[e.g.][]{Kliem:Toeroek:2006, Isenberg:Forbes:2007}.
In this case the smaller streamer field 
is less confining and the flux rope is able to reach the (lower) critical height 
for the torus instability to set in first.

\begin{figure}[htb!]
\centering
\includegraphics[width=0.8\textwidth]{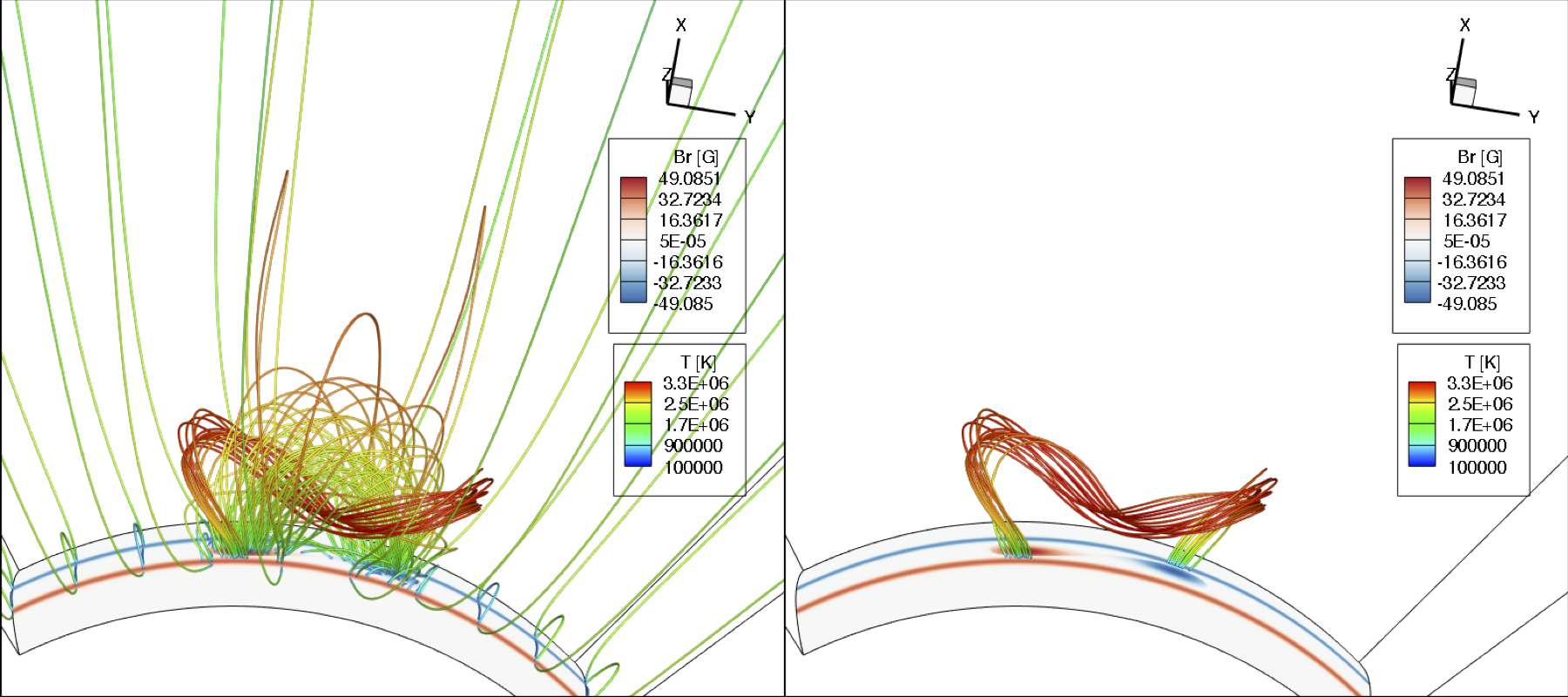}
\caption{The left panel shows the same
field lines as those in Figure \ref{fig:fdl3d_NS-S}(c), but colored in
temperature and with the addition of the field lines in the hot channel.
The right panel shows the hot channel field lines alone.
An animation of a rotating view of the field lines is available in the online
version.}
\label{fig:hotcorefdl_NS-S}
\end{figure}

The quasi-static rise of the flux rope after the
emergence is stopped is due to the tether-cutting reconnection in a current sheet
that forms underlying the flux rope, similar to what was found in \citet{Fan:2010}.
The reconnection continually add ``detached'' flux to the flux rope as
described in \citet{Fan:2010}, reducing its
anchoring and allowing it to rise quasi-statically to the critical height for the
onset of the torus instability.
The thermal signature of the tether-cutting reconnection is the formation of
a hot channel threading under the flux rope, containing heated twisted flux
added to the flux rope, as represented by the hot field lines shown in
Figure \ref{fig:hotcorefdl_NS-S}.
The left panel of the figure shows the same
field lines as those in Figure \ref{fig:fdl3d_NS-S}(c), but colored in
temperature and with the addition of the field lines in the hot channel,
and the right panel shows the hot channel field lines alone. An animation of
a rotating view of the field lines in both panels is also available in the
online version.
Note the hot channel field lines are rooted in the arcade field bands, and
therefore represent twisted flux newly added to the flux rope as a result
of multiple reconnections of the original arcade field lines with the flux
rope field lines.  These heated field lines which display a sigmoid shape
may correspond to the hot channel observed before and during CMEs 
by SDO/AIA described in \citet{Zhang:etal:2012} and \citet{Cheng:etal:2013}.
The temperature reached by the hot channel in our simulation is
about $3.3$ MK, significantly lower than the observed case, which is reported to
be as high as $10$ MK. However, this may be due to a much weaker field strength of
the flux rope and confining helmet field considered in our simulation.
Because of the significantly lower temperature of $3.3$ MK,
which is not picked up by the hot peaks of the temperature responses of
AIA 131 {\AA} channel (about 11 MK) and AIA 94 {\AA} channel (about 7 MK),
and in fact, it is at the
local valleys of these two channels' temperature responses, 
we do not see the sigmoid brightening corresponding to the hot channel 
field lines in the modeled emission (not shown here) in these channels.

\begin{figure}[hbt!]
\centering
\includegraphics[width=0.5\textwidth]{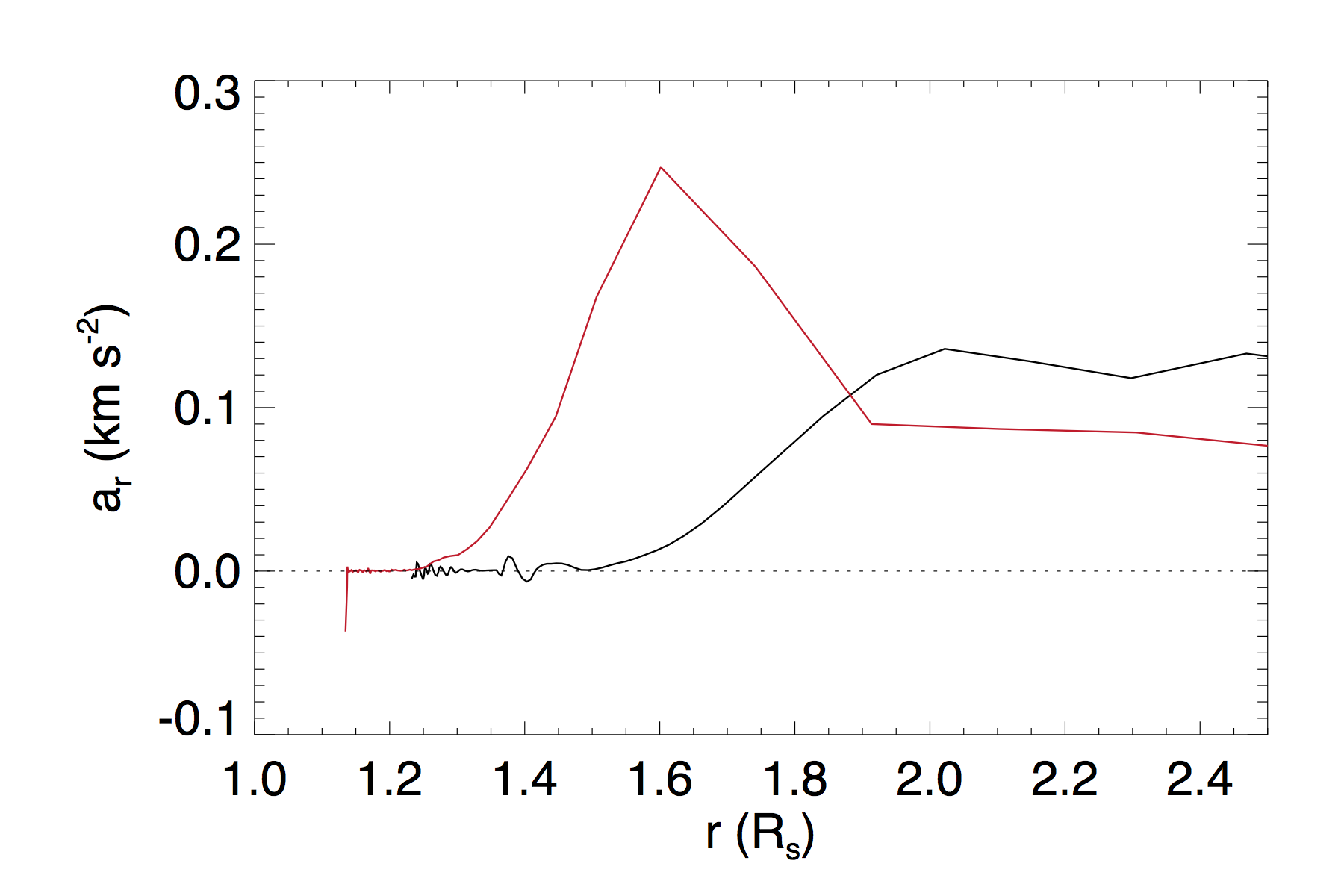}
\includegraphics[width=0.5\textwidth]{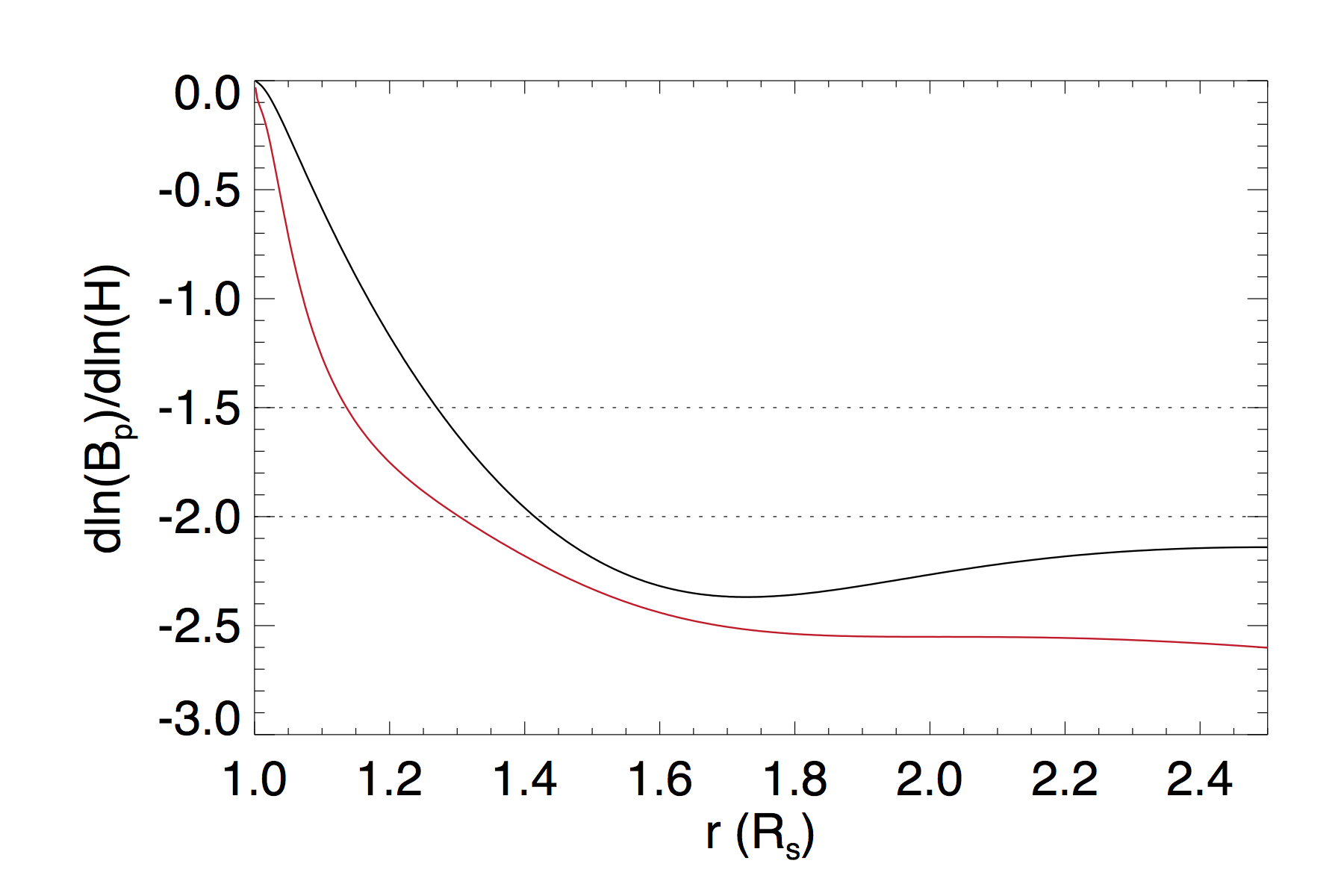}
\caption{(top panel) Acceleration at the apex of the axial field line 
as a function of its height position for the WS-L (black curve) and
NS-S (red curve) cases, and (bottom panel) the decay rate
with height of the corresponding potential field for the WS-L (black curve)
and NS-S (red curve) cases when the emergence is stopped (and hence no
more change of the lower boundary normal flux distribution and the
corresponding potential field).}
\label{fig:ar_vs_r}
\end{figure}

The top panel of Figure \ref{fig:ar_vs_r} shows the radial acceleration of the
tracked Lagrangian element at the apex of the flux rope axial field line as a
function of its height, comparing the NS-S case (red curve) and the WS-L case
(black curve).
It can be seen that the monotonic and significant acceleration for the dynamic
eruption sets in at a much lower height for the NS-S case
compared to the WS-L case.
The lower panel of Figure \ref{fig:ar_vs_r} shows the decay rate
$d ln (B_p ) / d ln (H)$ of the corresponding potential field $B_p$
with height above the solar surface $H$ along the vertical line through
the center of the apex cross-section of the flux rope, for the WS-L (black curve)
and NS-S (red curve) cases, when the emergence is stopped (and hence no more
change of the lower boundary normal flux distribution and the corresponding
potential field afterwards).
The decline with height of the potential field for the narrow streamer case
NS-S is significantly steeper compared to that for the wide streamer case WS-L,
explaining why for the NS-S case the torus instability sets in first at
a lower height. From the two panels of Figure \ref{fig:ar_vs_r}
we see that for the NS-S case the onset of significant acceleration takes
place when the tracked apex of the axial field line reaches about
$1.25 R_{\odot}$ at which the field decay rate is about -1.9, within
the range of critical decline rates (about $-1.$ to $-2.$) for the
onset of the torus instability obtained from several theoretical
calculations with simplified current loop models
\citep[e.g.][]{Titov:Demoulin:1999, Kliem:Toeroek:2006,
Demoulin:Aulanier:2010}.
For the WS-L case on the other hand, the onset of significant acceleration
of the flux rope takes place at a much higher height at about $1.55 R_{\odot}$
where the field decay rate is about $-2.3$, more than the nominal
range of the critical decay rate for the onset of the torus
instability.  But here the flux rope has already become significantly
kinked due to the onset of the kink instability first and its final
loss of equilibrium and eruption cannot be simply described by the
onset of the torus instability assuming a simple current path.
However the need for a sufficiently steep spatial decline of the
background potential field in order to achieve an ejective eruption
of the flux rope through the onset if the kink instability has
also been found in previous MHD simulations of kink unstable flux
ropes \citep[e.g.][]{Toeroek:Kliem:2005}.

\subsection{Formation and eruption of prominence}
\begin{figure}[th!]
\centering
\includegraphics[width=1.\textwidth]{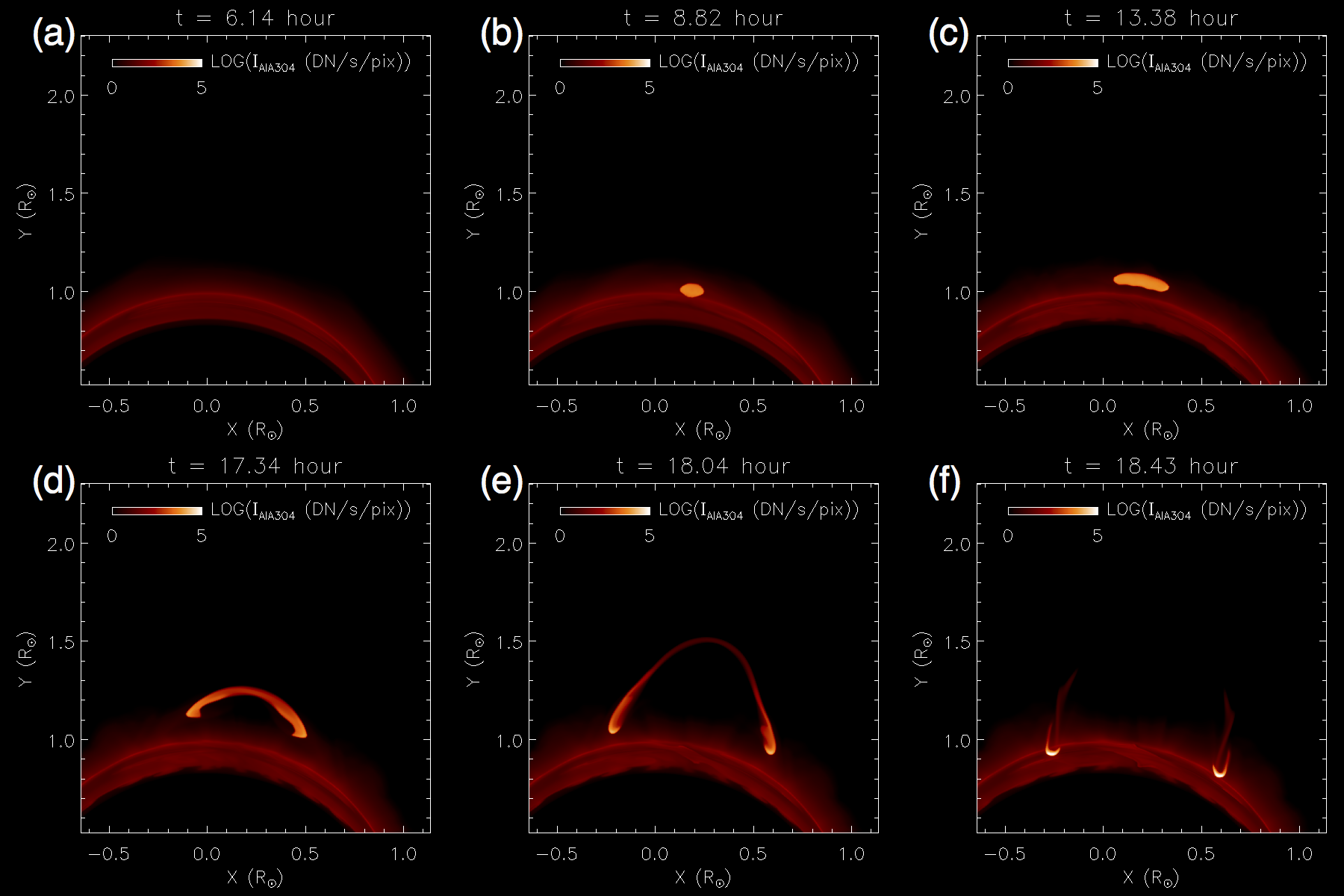}
\caption{Synthetic SDO/AIA 304 {\AA} channel emission images as viewed
from the same line of sight as that of panels (a)-(f) of Figure \ref{fig:fdl3d_WS-L} at
the same time instants, computed from the simulation case WS-L.
A corresponding movie
showing the evolution of the synthetic AIA 304 emission from $t=7.93$ hour to
$t=18.4$ hour is also
available in the online version.}
\label{fig:aia304_WS-L}
\end{figure}

The explicit inclusion of the optically thin radiative loss term 
(eq. \ref{eq:radloss}) in the energy equation provides the driver
for the development of radiative (thermal) instability or
non-equilibrium \citep[e.g][]{Priest:2014} which
allow the formation of cool dense prominence plasma in the hot rarefied corona.
As is described in \citet[][section 11.6.3 and references therein]{Priest:2014}
the physical form of the optically thin radiative loss term
implies that if a coronal plasma cools locally, the radiative loss increases
further leading to a run-away cooling.  This is true even if the cooling
function $\Lambda$ is constant with $T$ (let alone increasing with
decreasing T at certain temperature ranges), because density $N$ would
increase with a temperature decrease assuming pressure unchanged at
the perturbation.  This would further enhance the cooling because of the $N^2$
dependence on the radiative loss.  The instability can be suppressed by
thermal conduction but only if the length scale of the perturbation is
not too long.  Thus for sufficiently long coronal loops, the radiative
instability or non-equilibrium will develop.
We find in both the WS-L and WS-M cases, cool prominence condensations
with temperature as low as $7.3 \times 10^4$ K and density as high
as $5.6 \times 10^{9} {\rm cm}^{-3}$ develop in the corona in
the dips of the flux rope field lines.
Figure \ref{fig:aia304_WS-L} shows the synthetic SDO/AIA 304 {\AA}
channel emission,
computed by integrating along the line-of-sight
(that is the same as that for the view of panels (a)-(f) of Figure \ref{fig:fdl3d_WS-L})
through the simulation domain:
\begin{equation}
I_{\rm AIA304} = \int n_e^2 (l) \, f_{\rm AIA304} (T(l)) \, dl,
\end{equation}
where $l$ denotes the length along the line of sight through the simulation
domain, $I_{\rm AIA304}$ denotes the emission intensity at each pixel in
units of DN/s/pixel (shown in LOG scale in the images),
$n_e$ is the electron number density, and
$f_{\rm AIA304} (T)$ is the temperature response function that takes into
account the atomic physics and the properties of the AIA 304 {\AA} filter.
We have obtained the temperature dependent function $f_{\rm AIA304} (T)$ using
the SolarSoft routine {\tt get\_aia\_response.pro}.
The response function $f_{\rm AIA304} (T)$ peaks at the temperature of
about $8 \times 10^4$ K.
A movie of the evolution of the synthetic AIA 304 emission is also
available in the online version of the paper.
We see from Figure \ref{fig:aia304_WS-L} and the movie that prominence
plasma with temperature around $8 \times 10^4$ K, forms suspended in the
much hotter corona. It lengthens and develops into a suspended loop-like
structure during the slow quasi-static rise phase.
At the onset of the dynamic eruption of the flux rope, the prominence loop
also erupts with its apex rises upward while also showing substantial
draining and falling of the prominence plasma at the legs of the prominence
loop.
Such eruption morphology of a loop like structure with substantial draining
at the legs of the loop is often observed in prominence eruptions.
Although the flux rope field lines, especially those belonging to the
original inner flux surfaces (e.g. the blue field lines in
Figure \ref{fig:fdl3d_WS-L}), become significantly kinked
through the slow rise phase and the onset of
the eruption (with a rotation of $\sim 90^{\circ}$),
the prominence appears a loop like structure without showing
significant kinking when viewed from the same perspective as the flux rope.

\begin{figure}[ht!]
\centering
\includegraphics[width=0.85\textwidth]{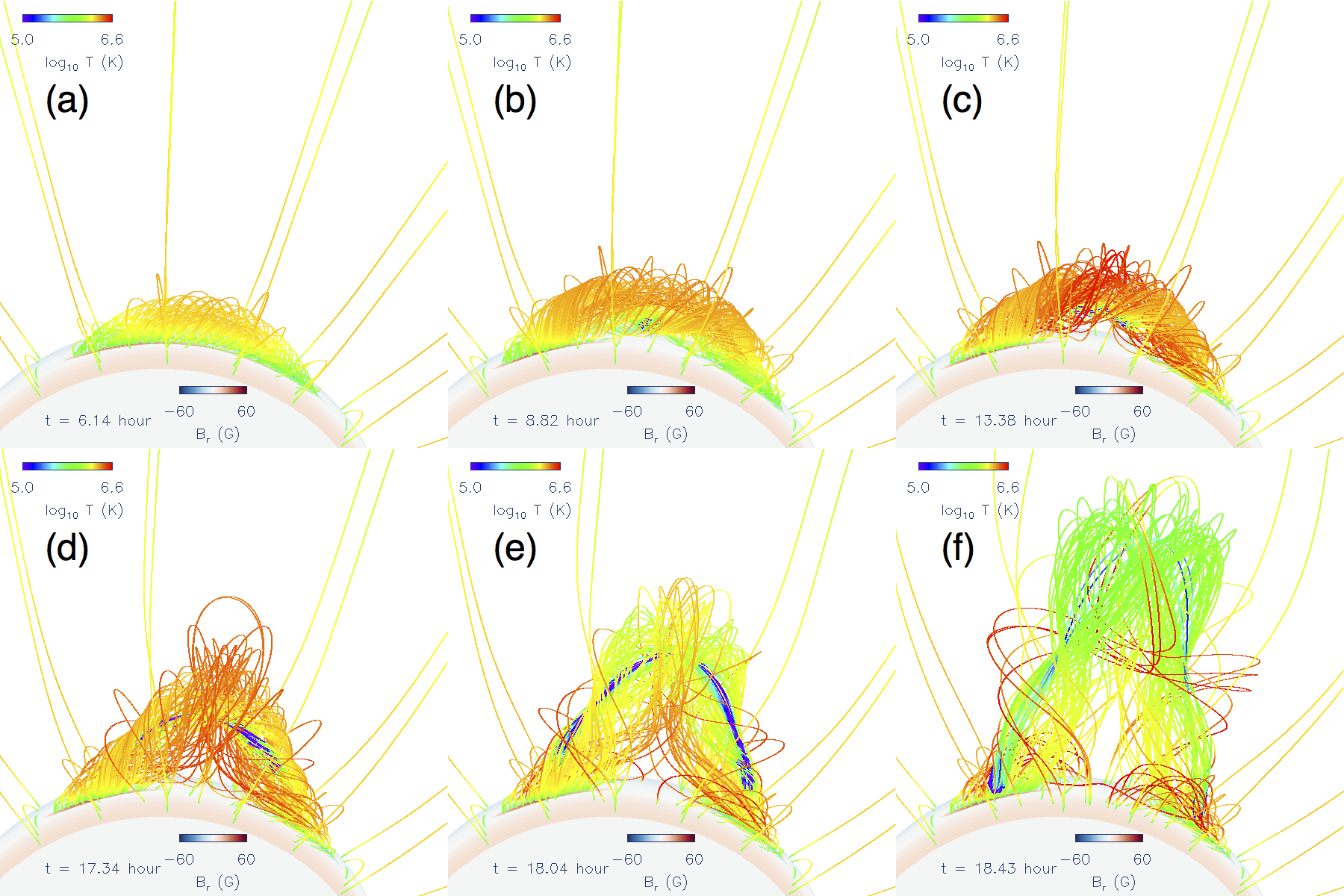}
\caption{Same as panels (a)-(f) of Figure \ref{fig:fdl3d_WS-L} but with the field lines
colored by the temperature, instead of based on the original flux
surfaces of the subsurface torus. Also additional prominence carrying field
lines are added by tracing field lines from grid points evenly sampled in
the region where temperature is below $10^5$ K. All images are viewed from
the same perspective as the AIA 304 images in Figure \ref{fig:aia304_WS-L}.}
\label{fig:combcfdl3d_WS-L}
\end{figure}
\begin{figure}[ht!]
\centering
\includegraphics[width=0.8\textwidth]{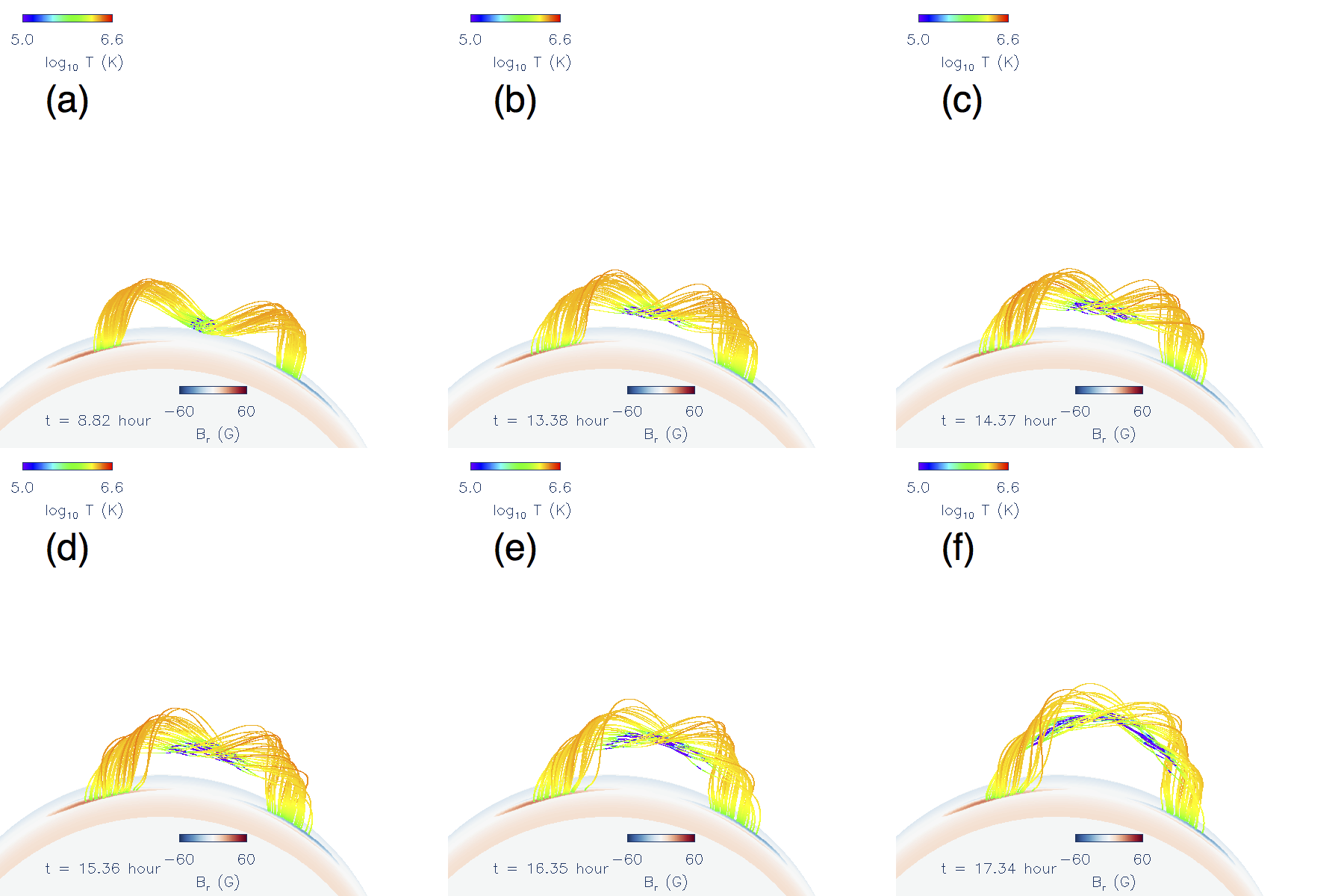}
\caption{Prominence carrying field lines colored in temperature during the
quasi-static rise phase. Field lines are
traced from grid points evenly sampled in the region where temperature is
below $10^5$ K. Note that panels (a), (b), and (f) here correspond to the
same time instants and the view as panels
(b), (c) and (d) of Figure \ref{fig:combcfdl3d_WS-L}, respectively, but
showing only the prominence carrying field lines by themselves.}
\label{fig:promfdl3d_side_WS-L}
\end{figure}
\begin{figure}[hb!]
\centering
\includegraphics[width=0.8\textwidth]{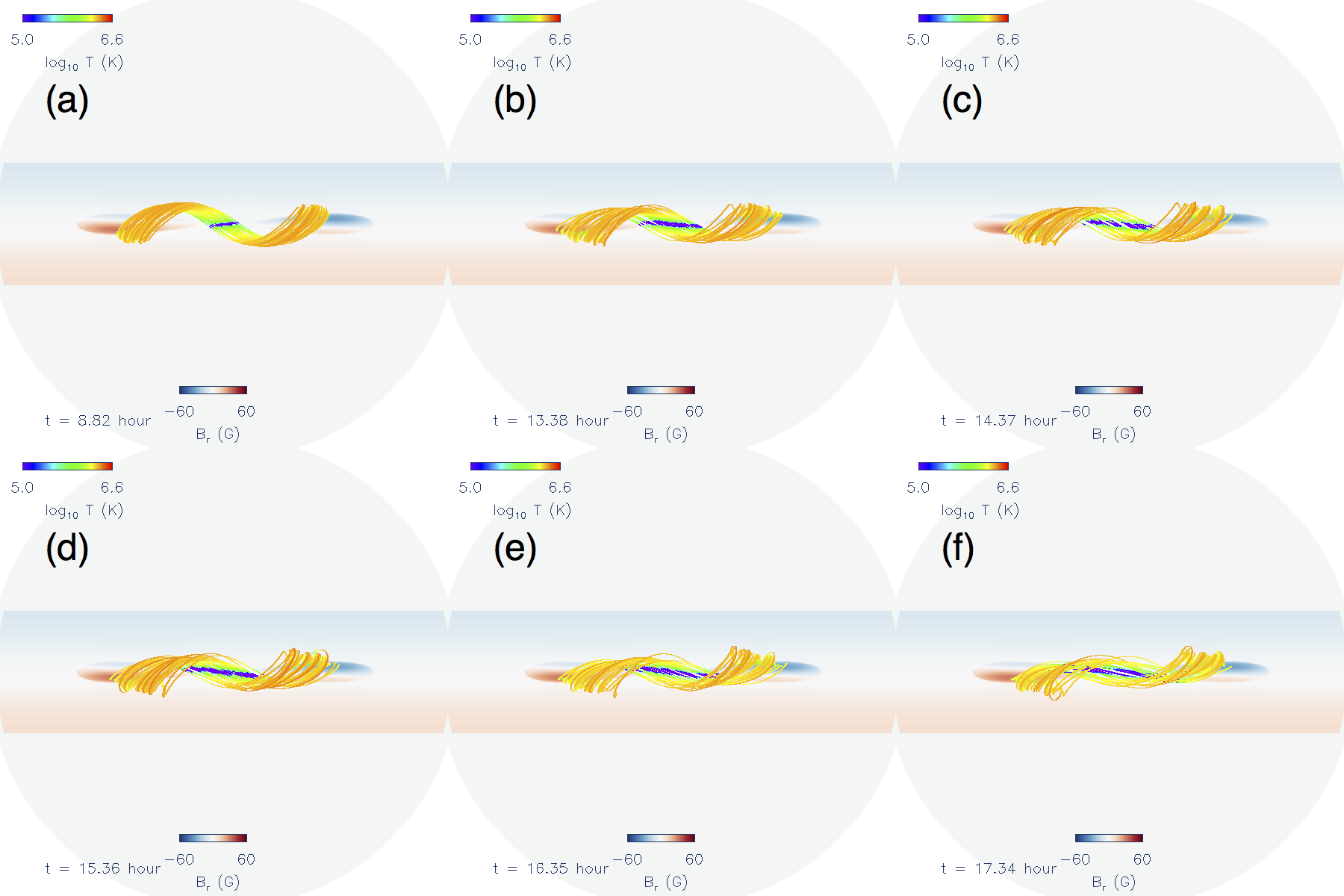}
\caption{Same as Figure \ref{fig:promfdl3d_side_WS-L}, but as viewed from the
top}
\label{fig:promfdl3d_top_WS-L}
\end{figure}

Figure \ref{fig:combcfdl3d_WS-L} shows the same snapshots of the 3D field
lines as those shown in Figure \ref{fig:fdl3d_WS-L} but with the field
lines colored by the temperature (instead of based on the original flux
surfaces) and also with additional prominence carrying field lines that
are traced from grid points sampled in the region where the
temperature is below $10^5$ K. The dark blue field line segments are
where prominence plasma is located, which are somewhat hard to see in
panels (b) and (c), as they are still small and
partially obscured by other field lines.
The location of the prominence condensations is more clearly seen in
Figure \ref{fig:promfdl3d_side_WS-L},
which show the prominence carrying field lines alone, during
the quasi-static rise phase.
Comparing to the AIA 304 images at the corresponding time instants
in Figure \ref{fig:aia304_WS-L}, we see that the prominence
condensations begin to form in the field line dips in
the lower middle part of the flux rope (see panels (b) and
(c) of Figure \ref{fig:combcfdl3d_WS-L} or more clearly panels (a) and (b) of
Figure \ref{fig:promfdl3d_side_WS-L}). The prominence continues to lengthen
during the quasi-static rise phase (panels (c), (d), (e), and (f) in Figure
\ref{fig:promfdl3d_side_WS-L}) and later erupts into
a loop like structure (see the blue field line segments in panels (d),
(e) and (f) of Figure \ref{fig:combcfdl3d_WS-L}).
Figure \ref{fig:promfdl3d_top_WS-L} shows the top view of the prominence
carrying field lines as those in Figure \ref{fig:promfdl3d_side_WS-L}
during the quasi-static rise phase.  It can be seen that as soon as
the prominence condensations develop into an elongated filament (the
dark blue segments in panels (b), (c), (d), (e), and (f) in Figure
\ref{fig:promfdl3d_top_WS-L}), its apparent orientation makes a small
angle (about $30^{\circ}$) with the orientation of the magnetic field
lines. This is consistent with the magnetic field configuration of
solar prominences as inferred from spectropolarimetric observations
\citep[e.g.][]{leroy:etal:1983,Bommier:etal:1994,Orozco:etal:2014}

\begin{figure}[ht!]
\centering
\includegraphics[width=0.83\textwidth]{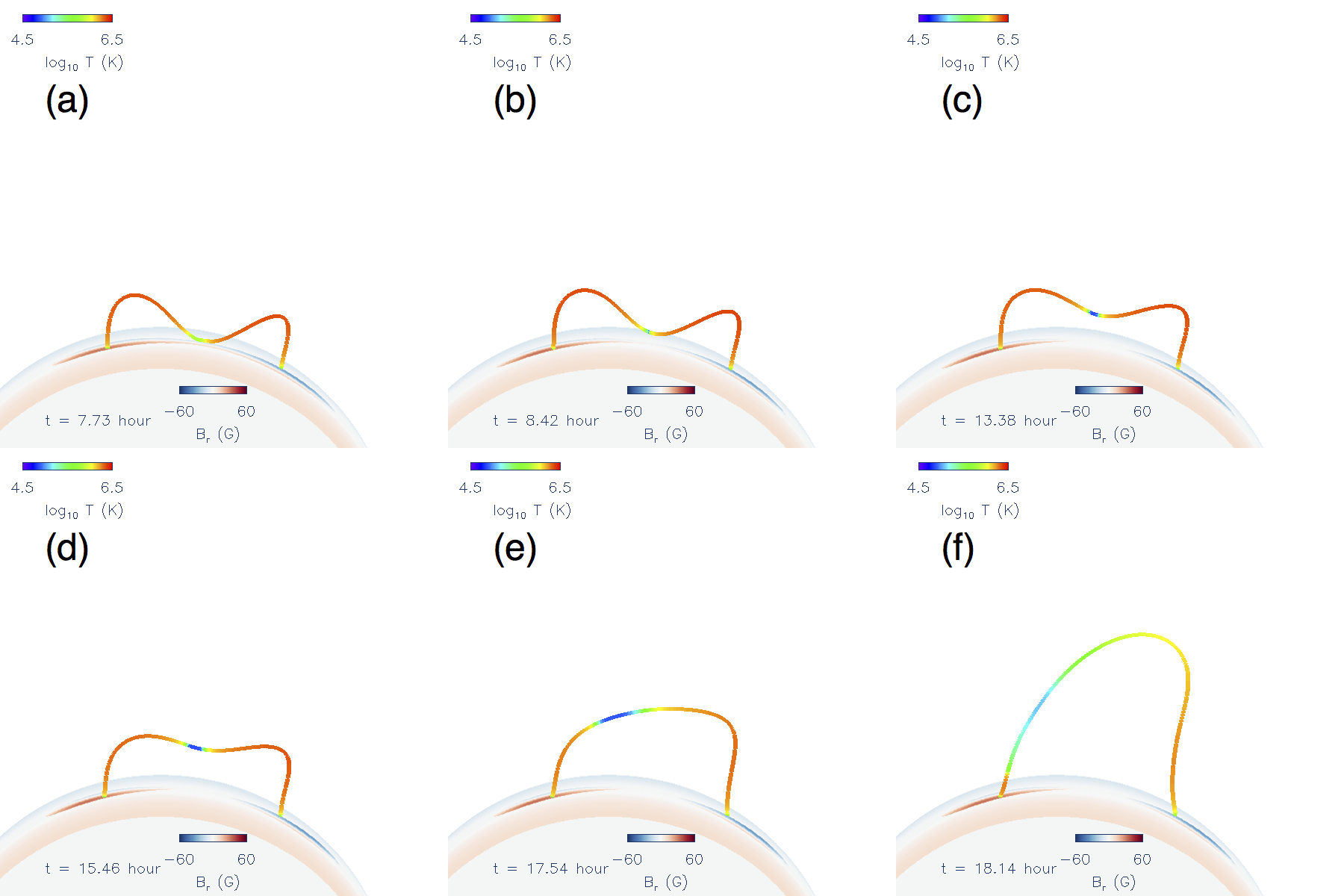}
\caption{The evolution of a prominence carrying field line colored with
temperature $T$.
The lower boundary sphere is colored with $B_r$.
A movie of this Figure showing the evolution from
$t=6.14$ hour to $t=18.4$ hour is available
in the online version.}
\label{fig:1fdlev_temp_WS-L}
\end{figure}

\begin{figure}[hb!]
\centering
\includegraphics[width=0.83\textwidth]{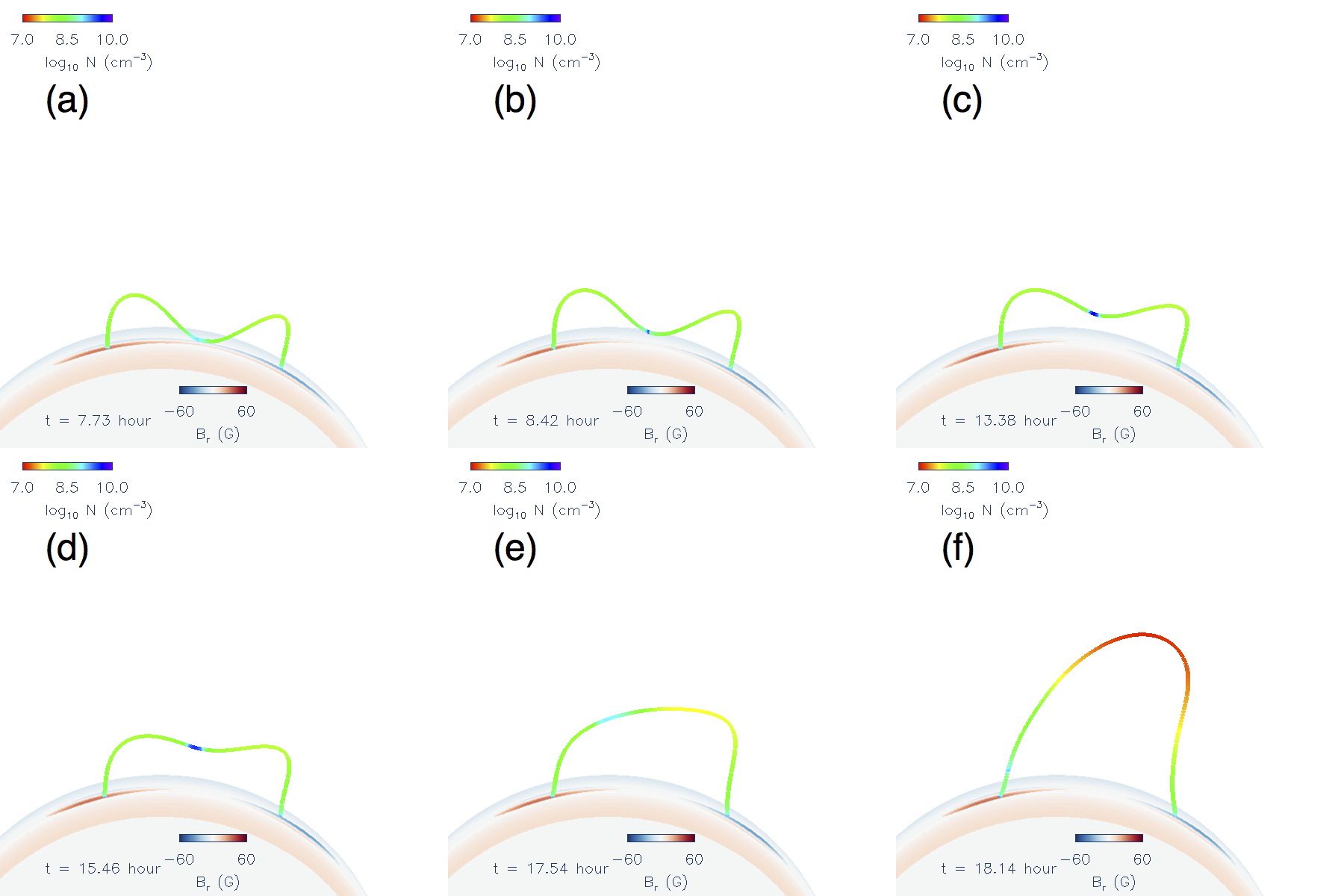}
\caption{The same as Fig. \ref{fig:1fdlev_temp_WS-L} but the field line is colored with density.
A movie of this Figure showing the evolution from
$t=6.14$ hour to $t=18.4$ hour is available
in the online version.}
\label{fig:1fdlev_den_WS-L}
\end{figure}

\begin{figure}[ht!]
\centering
\includegraphics[width=0.5\textwidth]{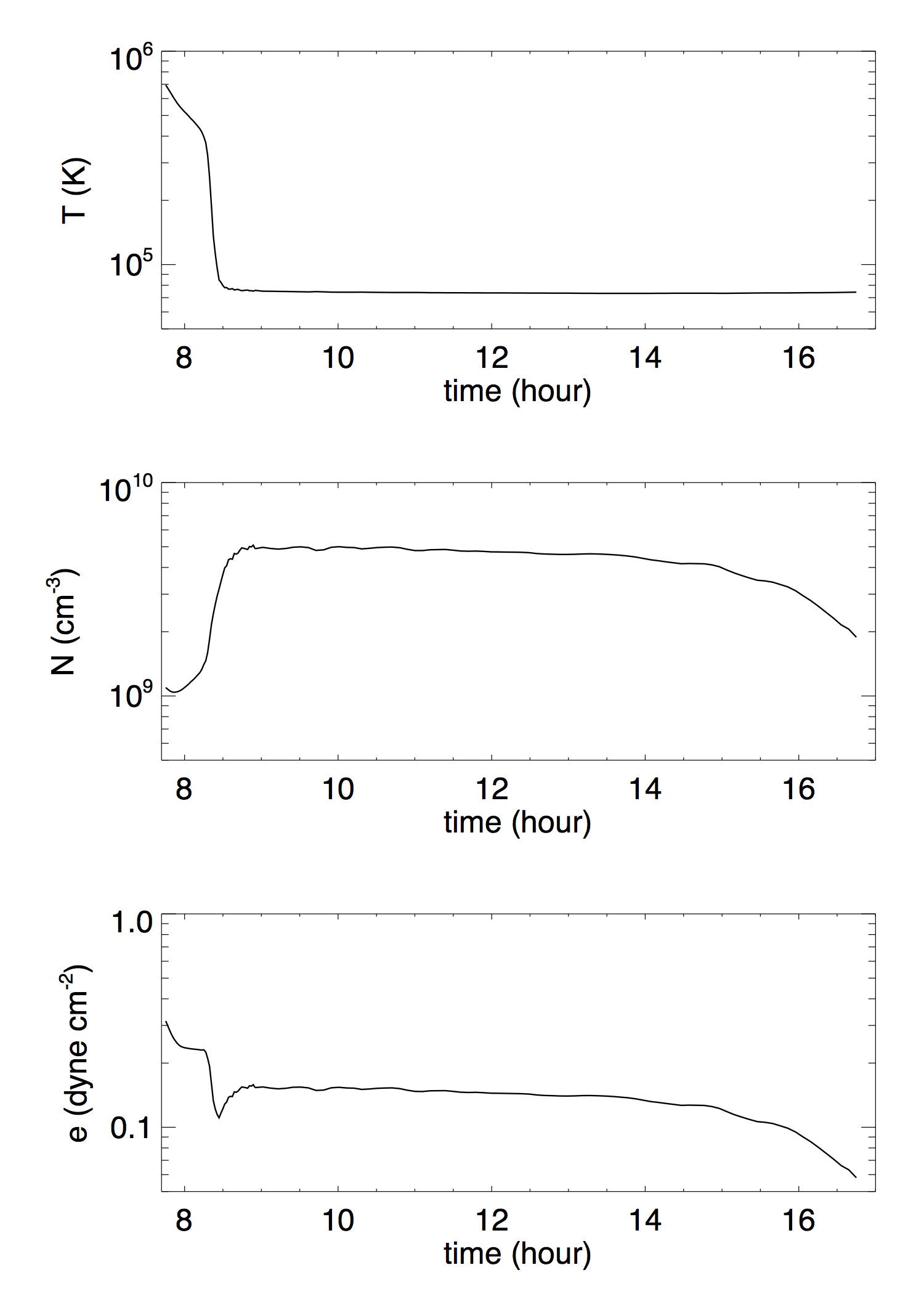}
\caption{The evolution of temperature $T$, density $N$, and internal energy
$e$ at the center of the dip of
the tracked field line shown in Figures \ref{fig:1fdlev_temp_WS-L} and
\ref{fig:1fdlev_den_WS-L}.}
\label{fig:trackdip_WS-L}
\end{figure}

Figure \ref{fig:1fdlev_temp_WS-L} and Figure \ref{fig:1fdlev_den_WS-L}
are snapshots showing the evolution of a tracked prominence field line
colored with temperature and density, respectively, along the field line.
Movies showing the evolution corresponding to each of these two figures are also
available in the online paper.
Figure \ref{fig:trackdip_WS-L} shows the evolution of temperature, density, and internal
energy at the center of the dip of the tracked field line shown in
Figures \ref{fig:1fdlev_temp_WS-L} and
\ref{fig:1fdlev_den_WS-L} from a time soon after the emergence of
the dip to about the time it disappears due to the onset of eruption.
\begin{figure}[ht!]
\centering
\includegraphics[width=0.5\textwidth]{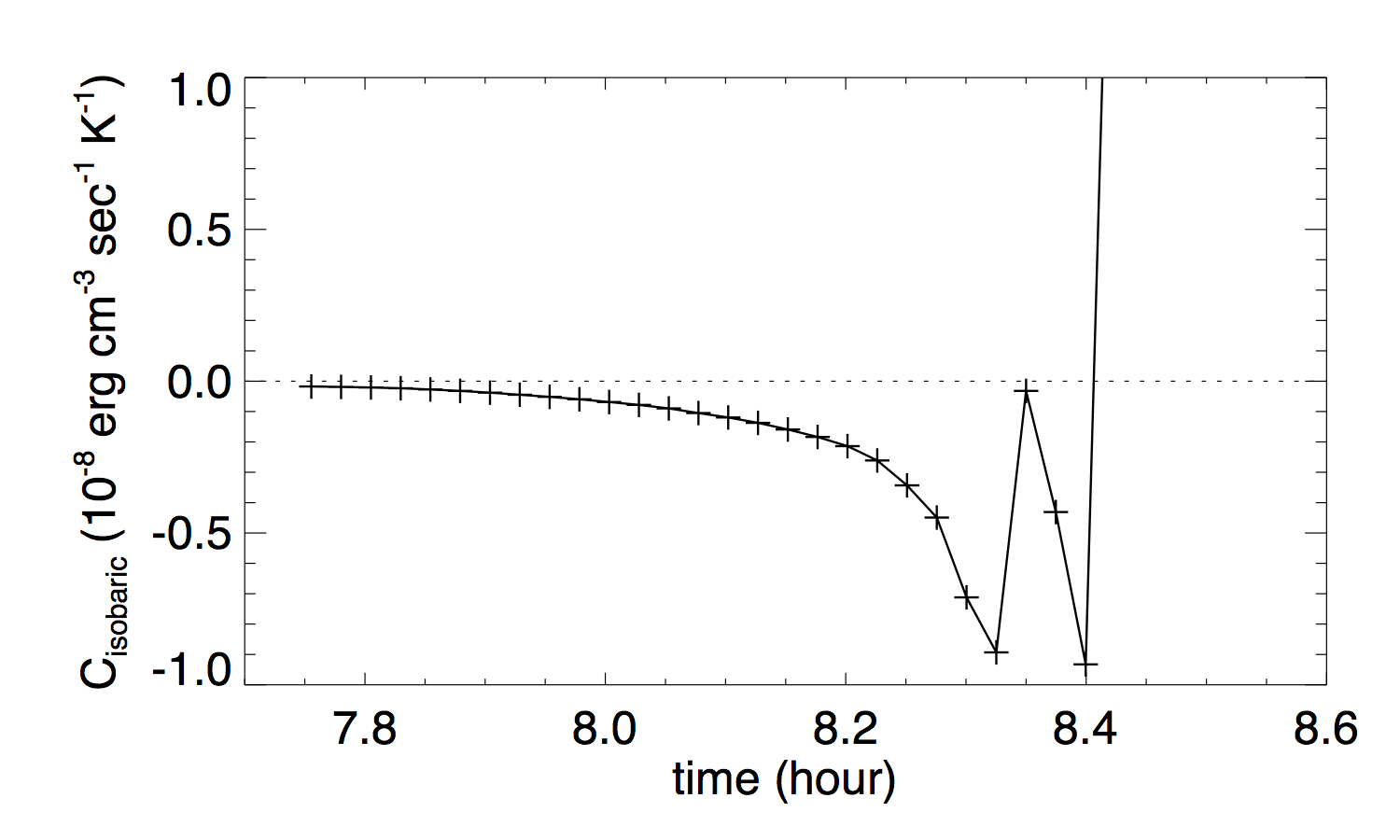}
\caption{Evaluation of $C_{\rm isobaric}$ for the criterion for isobaric thermal
instability at the center of the dip (see text).}
\label{fig:cisob}
\end{figure}
We can see that soon after the emergence of the dip (Figures
\ref{fig:1fdlev_temp_WS-L}(a) and \ref{fig:1fdlev_den_WS-L}(a)),
the plasma in the dip
has a coronal temperature of about $7 \times 10^5$ K and a density of about
$1.1 \times 10^9 {\rm cm}^{-3}$, and it is already not in thermal equilibrium, showing a
cooling with decrease in $T$ and $e$ (see top and bottom panels of Figure \ref{fig:trackdip_WS-L} at
about $t=7.7$ hour). The density shows a brief initial decrease as an initial dynamic
adjustment due to the rise of the dip and then shows a steady increase (middle panel of
Figure \ref{fig:trackdip_WS-L} at about $t=7.7$ hour). At about $t=8.25$ hour, an even
faster phase of cooling sets in with sharper decrease of temperature and pressure, and
sharper increase of density, until about $t=8.6$ hour, when the dip settles into a
thermal equilibrium for the rest of the time of the quasi-static phase.
We have also evaluated the criterion for the isobaric thermal instability as given in
\citet[][and references therein]{Xia:etal:2011}:
\begin{equation}
C_{\rm isobaric} \equiv \rho \left( \frac{\partial L}{\partial T} \right)_{\rho}
-\frac{\rho^2}{T} \left( \frac{\partial L}{\partial \rho} \right)_{T} + k^2 \kappa < 0,
\label{eq:cisob}
\end{equation}
where $L \equiv (Q-H)/\rho$ with $Q$ and $H$ given by equations \ref{eq:radloss}
and \ref{eq:heatingfunc} respectively, $\kappa = \kappa_0 T^{5/2}$ is the thermal
conductivity, and $k$ is the wavenumber for the thermal perturbation.
Equation \ref{eq:cisob} is the criterion for the thermal instability assuming
fixed pressure (isobaric), which is suitable under the condition that hydrostatic balance
is maintained. We evaluate $C_{\rm isobaric}$ at the dip, where we use the
width of the dip as the half wavelength for $k$, and estimate the width as the distance
along the field line between the two points where $T$ has risen to twice the value
at the bottom of the dip. We assume $H$ is unchanged since it is only a
function of height.
The result of $C_{\rm isobaric}$ is shown in Figure \ref{fig:cisob}, where we find
that the thermal instability criterion is met soon after the emergence from
$t=7.7$ hour until about $t = 8.4$, when $T$ at the dip has dropped to about
$T = 7.3 \times 10^4$ K at which the steep decline of cooling function with
decreasing temperature provides the strong stabilizing effect to suppress the
instability.  Thus the plasma at the dip is undergoing thermal non-equilibrium
soon after the emergence and does not find an equilibrium until it cools down
to about $7 \times 10^4$ K.
The sharper decrease of $T$ that sets in at about $t=8.25$ hour is due to
the enhanced cooling as $\Lambda$ increases sharply with decreasing $T$ at $T$
$\sim 4 \times 10^5$ K (see Figure \ref{fig:radloss}) compounded by the
increase in density. We also find that the
cooling time scale (estimated from $e/Q$ at the dip) decreases to become 
comparable to
the sound crossing time of the width of the dip at about this time
($t = 8.25$ hour) which may be the cause of the onset of the sharper drop
in $e$ or pressure, because the hydrostatic balance is no longer well maintained.
Through the course of the thermal non-equilibrium, we find that the
temperature at the dip drops from about $7 \times 10^5$ K to
about $T = 7.3 \times 10^4$ K, and remains there for the rest of
the quasi-static rise phase (Figure \ref{fig:trackdip_WS-L} and panels (b), (c), (d)
of Figure \ref{fig:1fdlev_temp_WS-L}).
The density increases from about $1. \times 10^9 {\rm cm}^{-3}$
to a peak value of about $5 \times 10^9 {\rm cm}^{-3}$, and remains above
about $3 \times 10^9 {\rm cm}^{-3}$
while the dip is present (Figure \ref{fig:trackdip_WS-L} and
panels (b), (c), and (d) of Figures \ref{fig:1fdlev_den_WS-L}).
When the dip disappears due to the onset of eruption the density
drastically reduces and part of the condensation drains down along the
left leg of the field line while part of the mass erupts with the top
portion of the loop (panels (e) and (f) of Figures \ref{fig:1fdlev_temp_WS-L}
and \ref{fig:1fdlev_den_WS-L}).

In our simulation we find that the peak density for the prominence condensation
form in the coronal domain reaches about $5.6 \times 10^9 {\rm cm}^{-3}$, about 30
times the density of the surrounding coronal plasma.
In the solar corona, the density ratio between the prominence and the surrounding
corona can be more than 100 \citep[e.g. see review in][]{Priest:2014}.
The reason that our simulation of the prominence condensation does not reach the
observed density is most likely due to the imposed suppression of the radiative
cooling for $T \leq 7 \times 10^4$ K as shown in Figure \ref{fig:radloss}, which
prevents the temperature at the prominence dip from going below
$7 \times 10^4$ K. This prevents the pressure at the dip from dropping lower
to draw more prominence mass from the field line foot points. It also limits
how high the density can be increased for the same pressure at the dip.
Furthermore, the way the lower boundary condition is imposed,
where the radial velocity needs to accelerate from nearly zero from
the lower boundary mass reservoir given the specified pressure
may constrain the mass flow to the condensation
at the dips.

\begin{figure}[ht!]
\centering
\includegraphics[width=0.5\textwidth]{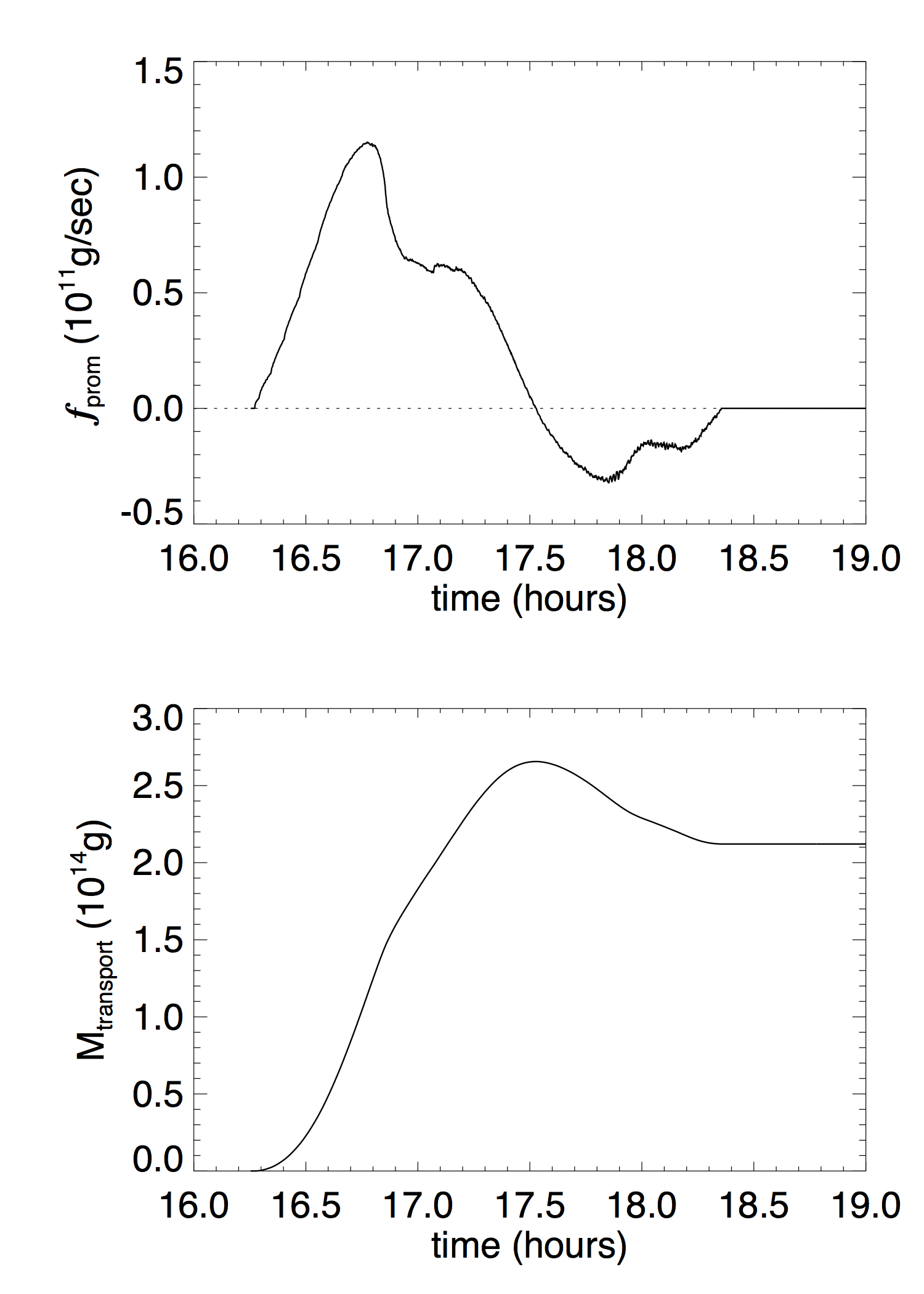}
\caption{(top) The prominence mass flux and (bottom) the
time integrated net prominence mass transported across the constant height
surface at $r=1.264 R_{\odot}$.}
\label{fig:promflux}
\end{figure}
It is difficult to estimate the percentage of prominence mass that is
ejected vs. drained down during the eruption because the temperature of
the plasma is changing.
Here we compute an integrated mass flux of cool prominence mass across
a certain constant height surface.
For this we consider cool prominence mass with temperature
below $10^5$ K. We find that the total cool prominence mass in the coronal
domain reaches a peak value of $ M_{\rm peak} = 5.46 \times 10^{14} g$
at time $t_{\rm peak}=16.3$ hours, with the prominence apex reaching
the height of $r_{\rm peak}=1.264 R_{\odot}$. Then starting from
$t_{\rm peak}$ we compute the prominence mass flux
$f_{\rm prom} = \int (\rho v_r )_{T < 10^5 K} dS$
through the constant height surface $S$ at $r_{\rm peak}$
and integrate this flux over time to obtain the net prominence mass
transport through S over time:
$M_{\rm transport} (t) = \int_{t_{\rm peak}}^{t} f_{\rm prom} dt $.
The result for $f_{\rm prom}$ and $M_{\rm transport}$ are shown in Figure
\ref{fig:promflux}.  We see an outward prominence flux $f_{\rm prom}$
from $t_{\rm peak} = 1.63$ hours to $t \approx 17.5$ hours, and an inward
flux $f_{\rm prom}$ from $t \approx 17.5$ hours to $t \approx 18.36$ hours,
after which there is no more flux of cool prominence material.  Thus the net
prominence mass transport $M_{\rm transport}$ out of the surface $S$ at
$r_{\rm peak}$ reaches a final value of about $2.1 \times 10^{14}$ g, which
is about 40\% of the peak prominence mass formed below $S$.
However there is uncertainty with this estimate because the prominence material
could have heat up to above $10^5$ K before rising through (or falling through)
the surface S, which would result in under estimating (over estimating) the
erupted prominence mass.  For this estimate we are
only taking into account net transport of mass that remains below $T$ of
$10^5$ K.

\begin{figure}[hbt!]
\centering
\includegraphics[width=0.5\textwidth]{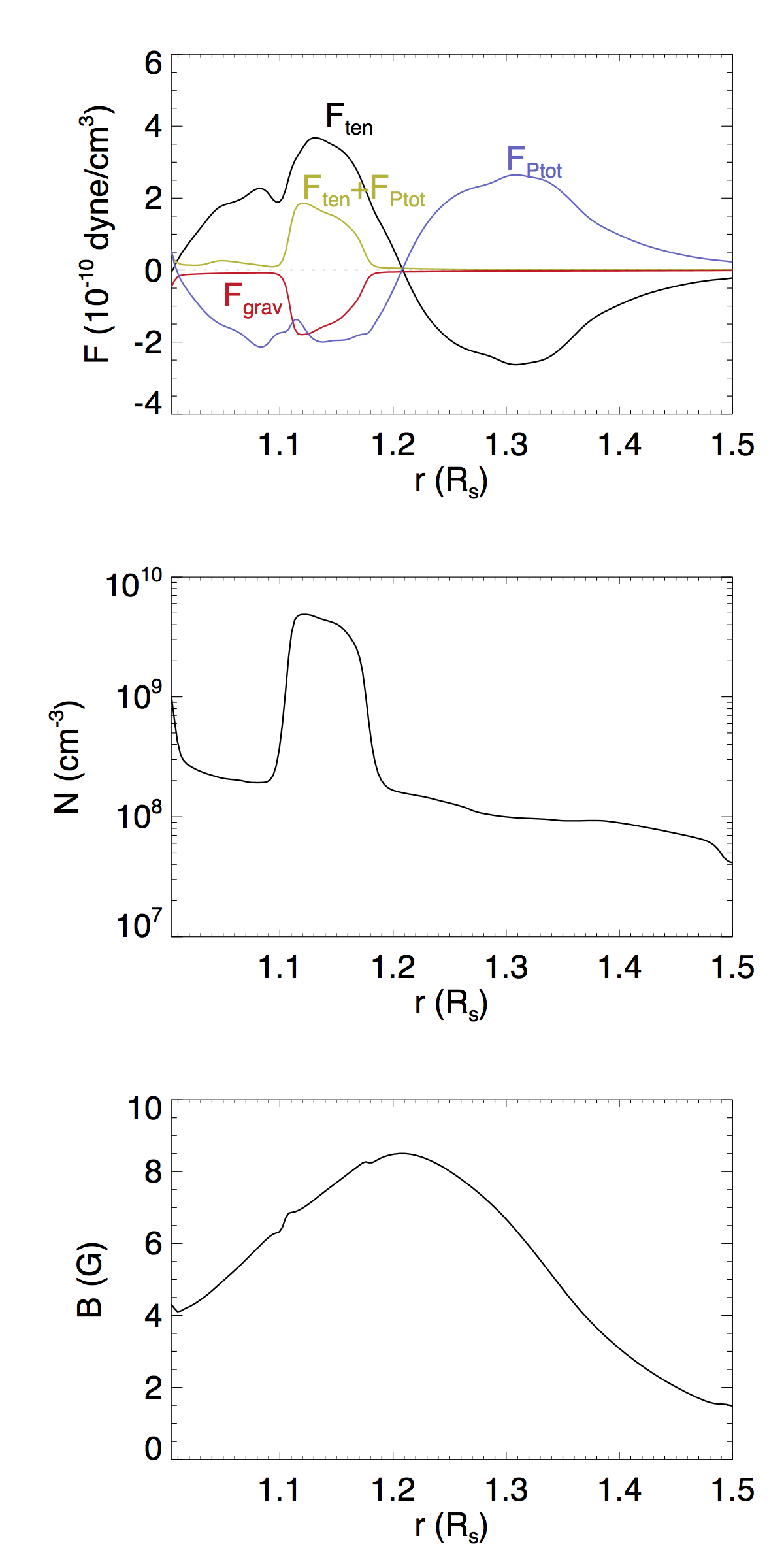}
\caption{Radial forces (top), density (middle), and total magnetic field strength
(bottom) along the central vertical line through the middle of the flux rope
shown in Figure \ref{fig:fdl3d_WS-L}(c). It also goes through the middle of
the prominence carrying field lines shown in
Figure \ref{fig:promfdl3d_side_WS-L}(b).
The radial forces shown in the top panel are the magnetic tension force
$F_{\rm ten}$ (black curve), the total pressure gradient force
$F_{\rm Ptot}$ (blue curve), where the
total pressure is mostly made up of the magnetic pressure, the
sum $F_{\rm ten}+F_{\rm Ptot}$ (green curve), which is approximately the
net Lorentz force, and the gravity force of the plasma
$F_{\rm grav}$ (red curve).}
\label{fig:forces_WS-L}
\end{figure}

We find that in the field line dips of the flux rope where the prominence
condensations form, the magnetic field becomes significantly non-force-free
because of the prominence weight.
Figure \ref{fig:forces_WS-L} shows the various radial forces (top),
density (middle), and total magnetic field strength
(bottom) along the central vertical line through the middle of the flux rope
shown in Figure \ref{fig:fdl3d_WS-L}(c). The vertical line also goes
through the middle of the prominence carrying field lines shown in
Figure \ref{fig:promfdl3d_side_WS-L}(b).
We can see from Figure \ref{fig:forces_WS-L} that in the height range
where the prominence has formed (indicated by the high density bump
in the middle panel), the downward gravity $F_{\rm grav}$ (red curve in
the upper panel) of the prominence plasma counteracts a significant
portion of the upward tension force $F_{\rm ten}$ (black curve in the
upper panel) of the magnetic field.
In the flux rope
the magnetic tension $F_{\rm ten}$ and the total pressure
gradient $F_{\rm Ptot}$ (where the total
pressure is largely made up of the magnetic pressure) are well balanced
and hence force-free outside of the prominence carrying region,
as can be seen by comparing
the black and the blue curves in the top panel. The sum
$F_{\rm ten}+F_{\rm Ptot}$ (green curve), which is approximately the net
Lorentz force, is nearly zero except in the prominence
carrying region, where it has a significant positive
net force to counteract the downward gravity force. 
Thus despite the fact that the plasma $\beta$ is low throughout the
flux rope (about 0.01 to 0.1 in the prominence region),
the magnetic field is significantly non-force-free in the region
of the prominence, where the prominence weight is
counteracting a major portion
of the upward magnetic tension, with the remainder balanced by the
downward magnetic pressure gradient force.
Note that the magnetic pressure gradient force (blue curve) is
downward in the prominence region to counteract partly the upward
tension, changing sign in the upper part of the flux rope where the
curvature changes sign.  Thus the magnetic field strength is increasing
with height in the prominence (see lower panel of Figure
\ref{fig:forces_WS-L}). This would be an observational signature
from prominence magnetic field measurement that indicates that the
prominence is associated with dipped or concave upturning magnetic
field lines.
Our result on the significantly non-force-free field in low-$\beta$ plasma
supporting the weight of the prominence in the field line dips is
consistent with the findings in previous
MHD models of prominences by \citep{Xia:etal:2012} and
\citet{Hillier:vanB:2013}.
   
\section{Discussion}
We have improved upon previous simulations of flux rope
destabilization and eruption by
using a helmet streamer pre-existing coronal field, and incorporating
a more sophisticated treatment of the thermodynamics that explicitly
include the non-adiabatic effects: an empirical coronal heating,
optically thin radiative losses, and field-aligned thermal conduction.
Depending on the size of the pre-existing streamer,
we find different scenarios
and mechanisms for the transition from quasi-equilibrium to ejective
eruption for the flux rope.
For a broad streamer with a slow decline of the magnetic field with
height, the flux rope is found to remain well confined until its
emerged twist is sufficiently high for the kink instability to set
in first. The kinked flux rope can still remain confined and goes
through a quasi-static, slow rise phase until its kinked apex 
reaches a certain height with sufficient decline of the confining
field where it develops a ``hernia-like'' ejective
eruption.
On the other hand with a narrow streamer with 
a steeper decline of the field with height, the flux rope can
erupt with a twist that is significantly below
the onset of the kink instability. 
It undergoes a quasi-static rise phase where tether-cutting
reconnections convert arcade flux into twisted flux of the flux
rope, reducing the confinement, and develops an ejective eruption
when it reaches the critical height for the onset of the torus instability.
The above mechanisms for the onset of eruption are in qualitatively
agreement with previous findings in simulations using a potential
pre-existing field and with simplified treatment of the thermodynamics
\citep[e.g.][]{Toeroek:Kliem:2005, Fan:Gibson:2007, Aulanier:etal:2010,
Fan:2010}.
Our simulations confirm that the fast decay with height
of the confining helmet magnetic field is a key factor for achieving an
ejective eruption of the underlying flux rope.
Our simulations also show that with the more realistic adiabatic 
index of $\gamma=5/3$, which produces a stronger adiabatic cooling
of the expanding erupting flux rope, and with an explicit treatment of
the heating and heat transport, the erupting flux rope is still able
to accelerate to a typical CME terminal speed ($\sim 600$ km/s in our
ejective eruption cases) in excess of the ambient
solar wind speed.

We have also achieved a simulation of the prominence eruption.
With the explicit inclusion of the optically thin radiative losses, we
found the formation of prominence condensations with a temperature
as low as $7.3 \times 10^4$ K and density as high
as $5.6 \times 10^{9} {\rm cm}^{-3}$
in the field line dips of the significantly twisted
flux ropes (in the WS-L and WS-M cases) during the quasi-static phase.
The prominence condensations are formed in the field line dips after
the emergence into the corona, as a result of the onset of radiative
instability or non-equilibrium.
The prominence condensations develop into an elongated structure
suspended in the corona as viewed in SDO/AIA 304 {\AA}
emission.
The elongated prominence structure makes an acute angle of about
$30^{\circ}$ with the orientation of the field lines supporting
the prominence, consistent with observations.
However, once formed due to runaway radiative cooling, the pressure
scale height of the coolest part of the prominence condensations (reaching
a minimum of 4.4 Mm) is only resolved by about 2 grid points given our
numerical resolution and hence not well resolved.
Thus their evolution is likely significantly impacted by numerical diffusion
and is probably reflecting an averaged collective motion of the
condensations.
We find that because of the weight of the prominence condensations
that formed, the prominence carrying field becomes significantly
non-force-free (despite being low plasma $\beta$) with a
significant fraction of the magnetic
tension force counteracting the gravity force of the prominence,
and with the remainder upward tension force of the concave
field lines balanced by a downward magnetic pressure gradient
force.
This confirms the previous findings by \citet{Xia:etal:2012}
and \citet{Hillier:vanB:2013}.
Thus the formation of the prominence may be playing a
significant role in increasing the confinement of the flux rope.

With the eruption of the flux rope in the WS-L case and the
disappearance of the dips, we find that the prominence 
plasma shows substantial draining along the legs of the
erupting field lines, developing into an erupting loop structure
as viewed in AIA 304 {\AA} emission.
The erupting prominence obtained here does not show a kinked
morphology, even though the flux rope becomes significantly kinked when
viewed from the same perspective.
We find that the cool prominence condensation (with $T < 10^5$ K)
reaches a peak mass of
about $5.46 \times 10^{14} g$ in the corona and estimate that
roughly 40\% of the cool prominence mass is transported out
with the eruption.
These results on the evolution of the prominence need
to be further investigated with higher resolution simulations
that better resolve the prominence condensations that develop.

In the case with the shorter, less twisted flux rope (NS-S case),
which eventually erupts due to the onset of the torus instability,
we do not find the formation of prominence condensations. Instead
we find the formation of a sigmoid-shaped hot channel which contains
heated, twisted flux added to the flux rope as a results of
tether-cutting reconnections during the quasi-static rise phase.
This thermal signature of tether-cutting reconnection may explain
the hot channels observed by SDO/AIA before and during the onset
of CMEs as described in e.g. \citet{Zhang:etal:2012}
and \citet{Cheng:etal:2013}.
However to quantitatively produce the observed temperature of the hot
channels (about 10 MK) so as to be able to model the bright emissions of
the hot channels seen in AIA 131 {\AA} and AIA 94 {\AA} images, simulations
with a significantly increased field strength for the flux rope and the confining
helmet streamer are needed.

\acknowledgments
I thank the anonymous referee for helpful comments that improved the paper.
I also thank Feng Chen and Sarah Gibson for reading the paper and helpful
comments.
This work is supported in part by the Air Force Office of Scientific Research
grant FA9550-15-1-0030 to NCAR. NCAR is sponsored by the National Science
Foundation. The numerical simulations were carried out on the Cheyenne
supercomputer at NWSC under the NCAR Strategic capability project NHAO0011 and
also on the DOD supercomputers Thunder at AFRL and Topaz at ERDC under the
project AFOSR4033B701.

\end{document}